Research article

# Interval type-2 fuzzy logic system based similarity evaluation for image steganography

Zubair Ashraf [a], Mukul Lata Roy [a], Pranab K. Muhuri [a,*], Q.M.Danish Lohani [b]

[a] Department of Computer Science, South Asian University, Akbar Bhavan, Chanakyapuri, New Delhi 110021, India
[b] Department of Mathematics, South Asian University, Akbar Bhavan, Chanakyapuri, New Delhi 110021, India

ARTICLE INFO

ABSTRACT

Similarity measure, also called information measure, is a concept used to distinguish different objects. It has been studied from different contexts by employing mathematical, psychological, and fuzzy approaches. Image steganography is the art of hiding secret data into an image in such a way that it cannot be detected by an intruder. In image steganography, hiding secret data in the plain or non-edge regions of the image is significant due to the high similarity and redundancy of the pixels in their neighborhood. However, the similarity measure of the neighboring pixels, i.e., their proximity in color space, is perceptual rather than mathematical. Thus, this paper proposes an interval type-2 fuzzy logic system (IT2 FLS) to determine the similarity between the neighboring pixels by involving an instinctive human perception through a rule-based approach. The pixels of the image having high similarity values, calculated using the proposed IT2 FLS similarity measure, are selected for embedding via the least significant bit (LSB) method. We term the proposed procedure of steganography as 'IT2 FLS-LSB method'. Moreover, we have developed two more methods, namely, type-1 fuzzy logic system based least significant bits (T1FLS-LSB) and Euclidean distance based similarity measures for least significant bit (SM-LSB) steganographic methods. Experimental simulations were conducted for a collection of images and quality index metrics, such as PSNR (peak signal-to-noise ratio), UQI (universal quality index), and SSIM (structural similarity measure) are used. All the three steganographic methods are applied on dataset and the quality metrics are calculated. The obtained stego images and results are shown and thoroughly compared to determine the efficacy of the IT2 FLS-LSB method. We have also demonstrated the high payload capacity of our proposed method. Finally, we have done a comparative analysis of the proposed approach with the existing well-known steganographic methods to show the effectiveness of our proposed steganographic method.

## 1. Introduction

Similarity describes the relationships between conceptual or perceptual entities (Li and Liu, 2015). It has been evolved from different senses (Demirci, 2007), as given below:

a) *Mathematical approach:* A number of mathematical distance functions have been developed to calculate the similarity between pairs of numerical data, e.g., Hamming distance, Euclidean distance, and so on.
b) *Psychological approach:* In terms of psychology, similarity has been studied by calculating the correlation between objects, human perceptual resemblance, and rule-like or theory-like semantic representations.
c) *Fuzzy approach:* The similarity is measured through fuzzy logic by simulating the human perception of similarity. It has the benefit of being able to handle non-numerical quantities, degrees of similarity, and perceptual reasoning occurring due to human judgments.

In the field of image processing, similarity measures for image pixels are typically evaluated using Euclidean distance in color space. However, Wuerger et al. (Wuerger et al., 1995) demonstrated that proximity in color space – something that is perceptual rather than mathematical – cannot be evaluated using this distance measure. In other words, the measure that we get using the Euclidean distance will not be adequate for judging the similarity of image pixels (Demirci, 2007). Moreover, since the similarity of pixels involves an instinctive approach rather than a numerical one, it becomes necessary to consider human perception in its measurement. Based on the fuzzy sets (Zadeh, 1965), fuzzy logic (FL) has concerns with the degree of membership in a set. It differs from conventional logic and is an essential tool to handle the problems in which imprecise information is associated with data. Prof. Zadeh himself was the first to offer a way to judge similarity in terms of FL in (Zadeh, 1971), where he extended the theories of relations and equivalence to adapt for problems where the parameters are not well-defined.

* Corresponding author.
*E-mail address:* pranabmuhuri@cs.sau.ac.in (P.K. Muhuri).

In the modeling of a real-life system, the incorporation of uncertainties plays an essential part. The uncertainties arise or exist because of lack of knowledge of system experts and incomplete information regarding the system inputs. Such causes impreciseness and ambiguity and inconsistency in the modeling of the system. Another important reason that causes the uncertainty in the system is the differences in the opinions of experts. In fuzzy sets or type-1 fuzzy sets (T1 FSs), the single expert opinion is incorporated with the membership function that signifies the degree of uncertainty. However, due to the conflicts in the opinions of various experts and various resources of uncertainty, T1 FSs are insufficient for modeling those systems. To overcome this, Prof. Zadeh himself proposed the concept of type-2 fuzzy sets (T2 FSs) (Zadeh, 1975). Further, Mendel et al. pioneered the interval type-2 fuzzy sets (IT2 FSs) to reduce the burden of computation with general T2 FS (Mendel and Wu, 2010). Since then, a significant number of applications have been developed using IT2 FSs, such as machine learning (John et al., 2000), image processing (Castillo et al., 2007), pattern recognition (Melin and Castillo, 2013), reliability engineering (Ashraf et al., 2014, 2015; Muhuri et al., 2017), real-time system (Muhuri et al., 2020a, b), similarity measures (Li et al., 2015), inventory system (Ashraf et al., 2017) etc.

Steganography is a technique of protecting information in such a way that it is not evident to unauthorized entities. Such a task has become increasingly more critical as the popularity of the internet as a medium of transmission increases. A steganographic method, therefore, aims to hide away secret data as unobtrusively as possible; this is typically done by fixing the data bits mentioned above deeply in another form of data. Both, the data to be embedded and the data needed for embedding, can be of several kinds: text, image, audio or video (Cheddad et al., 2010; Johnson and Katzenbeisser, 2000; Petitcolas et al., 1999). Steganographic methods consist of two parts: embedding and extraction. Embedding is the process of hiding the data in the pixels of the image, while extraction is the process of retrieving the hidden data from the image. The image used for embedding is called the cover image, and the output image after embedding is called the stego image. The overall objective of steganography – to hide the data so that it is indiscernible – can be achieved by ensuring a few things: the protected data inside the cover object must be transparently imperceptible; the method of embedding itself must be able to withstand various attacks; the stego object should have a high embedding capacity for hiding data bits; and the method used should be such that it can resist any damaging interference (Jafari et al., 2013; Jero et al., 2016; Kanan and Nazeri, 2014; Li et al., 2010; Subhedar and Mankar, 2014; Wu and Tsai, 2003).

In literature, there are several existing methods for steganography. These methods can be grouped based on their processing domain as follows: (1) spatial domain and (2) transform domain. The difference between these two domain techniques is that, in the spatial domain techniques, the embedding occurs directly in the pixels of the image, while in the transform domain techniques, the image is first transformed into the frequency domain and then the embedding is done. In image steganography, embedding in the plain or non-edge regions of the image has a significantly low effect on the human visual system. In other words, the pixel values in the plain (non-edge) regions of the image are similar, so they are good areas to embed. This is because the differences in the neighborhood pixel values of the plain region are very low (Melin et al., 2014). However, embedding in the textured region of the image is not advisable, since a slight change in the pixel values of that region can distort the image.

Additionally, the pixels, having integer values ranging between 0 and 255, are redundant to its neighbors and the levels of differences between them are almost negligible. Therefore, the corresponding differences based on numeric pixel values are not truly justified as similar or not similar. Thus, to perform image steganography in those regions, rule-based categorization is required to calculate the similarity value for any pair of pixels (Demirci, 2006). In other words, the perceptual reasoning of the similarity of the redundant pixels in the neighborhood should be assessed by human judgment (Hampton, 1998; Seaborn et al., 2005). Therefore, linguistic variables are the most appropriate for the modeling of perceptual similarity, and FSs have long been the best-recognized tool to represent these linguistic values.

One of the most fundamental issues in the linguistic variable representation by employing T1 FSs is the selection of the membership function (MF) that interprets the uncertainty in information. Additionally, there exists a need to justify the crispness of the membership value of the fuzzy set, as it seems to contradict the core idea of "fuzziness". The type-1 membership function (T1 MF) is mostly recommended by the system expert to signify the degree of belongingness to a T1 FS. However, when there is more than one expert involved to provide the opinions regarding the MFs, T1 FSs risk being unable to capture the different kinds of uncertainties. The above situation can be resolved by considering the union of all the T1 MFs representing the various expert opinions. This causes the generation of interval type-2 membership functions (IT2 MFs), leading to IT2 FSs (Ashraf et al., 2018a; Muhuri et al., 2017). Therefore, we propose that the linguistic terms be portrayed as IT2 FSs. The IT2 MFs of IT2 FSs are able to capture higher uncertainty compared to T1 FSs. In IT2 MFs, the Footprint of Uncertainty (FOU) depicts a degree of uncertainty caused by the differences of human judgment (or expert opinion) regarding the T1 MFs. The FOU offers more flexibility in the adjustment of the decisive parameters, i.e., the IT2 MFs of linguistic variables. As a result, the FOU of IT2 MFs are bounded by lower and upper membership functions representing the uncertainty of the linguistic terms . Hence, a significant improvement has been made in shifting from a type-1 fuzzy logic system (T1 FLS) to interval type-2 fuzzy logic system (IT2 FLS) by the researchers in recent years. IT2 FLSs have been effectively applied in the various applications of image processing systems such as classification (Majeed et al., 2018; Rubio et al., 2017), filtering (Singh et al., 2018), segmentation (Dhar and Kundu, 2019; Zhao et al., 2019), and edge detection (Castillo et al., 2017; Gonzalez and Melin, 2017; Gonzalez et al., 2016; Martínez et al., 2019; Melin et al., 2014).

In this paper, we propose a novel steganographic procedure using an IT2 FLS based similarity measure to measure the similarity between the pixels in a digital image. For doing so, a (3×3) window of pixels is selected to find the similarity of the central pixel to its neighboring pixels. We calculate the difference of all the three color components, i.e., red (R), green (G), and blue (B), which are assigned linguistic terms. The result from the IT2 FLS is a similarity matrix which contains values in the interval [0, 1] indicating the level of similarity of each pixel of the image to its neighbors. This similarity matrix is used to select pixels to perform the least significant bit (LSB) method for hiding the secret message. We have not found any previous work of similarity measure or steganography which has used an IT2 FLS. Cover images, namely, Lena, Baboon, Jet, Barbara, Boat, Peppers, Earth, House, Sailboat and Splash, have been chosen as the dataset, upon which the proposed steganographic method was applied to generate the corresponding stego images. Quality index metrics are used to assure the visual quality of the stego images. We have also demonstrated the high payload capacity of our proposed method.

The significant contributions of this paper are summarized as:

1) A new algorithm based on IT2 FLS was developed that calculates the perceptual similarity of an image.
2) Linguistic variables represent the color differences between pixels as {Low, Medium, High} and degree of similarity as {Not Similar, Slightly Similar, Moderately Similar, Quite Similar, Exactly Similar}. These variables are formulated as IT2 MFs in the IT2 FLS.

3) IT2 FLS takes the IT2 MFs of the linguistic variables as inputs and provides the output describing the similarity between a pair of pixels.
4) Mamdani inference engine using fuzzy-rules operates on the IT2 MFs to achieve the similarity. Consequently, the type-reduction and defuzzification generates the similarity value of a pixel in an image.
5) Similarity matrix depicting the similarity of each pixel in an image is evaluated based on a set of fuzzy-rules.
6) The embedding and extraction procedures of steganography are performed on the image using the least significant bit (LSB) method.
7) A pixel of an image is selected for embedding if the corresponding similarity value (in the similarity matrix) is greater than a particular threshold value, which indicates the degree of similarity to its neighbors.
8) To show the efficacy of our proposed method, type-1 fuzzy logic system (T1 FLS) based and Euclidean distance based similarity measures were also implemented to perform steganography.
9) The experimental simulations were performed on a dataset consisting of ten different cover images, and the corresponding stego images are obtained. The whole procedure with the three methods: (a) proposed, (b) T1 FLS based, (c) Euclidean distance based, was repeated for different numbers of bits ($k = \{1, 2, 3, 4\}$) hidden in the LSBs of the pixels and for different thresholds.
10) A thorough comparison has been done by investigating the perceptual transparency and visual quality of the stego images. These were measured by using three quality index metrics: peak signal-to-noise ratio (PSNR), structural similarity measure (SSIM), and universal quality index (UQI).

The rest of the paper is organized as follows: in Section 2, a literature survey is presented, detailing past works relevant to this paper. Mathematical definitions of T2 FSs, IT2 FSs, and IT2 FLSs are explained in Section 3. Steps of the proposed IT2 FLS based similarity measure technique and consequent steganographic method is elaborated in Section 4. Details of the experimental simulations and results are reported in Section 5, which also includes a comparative discussion. Finally, we conclude in Section 6.

## 2. Literature survey

Similarity measures played important role in many image processing applications such as edge detection (Demirci, 2007), filtering (Elmas et al., 2013), retrieval (Varish et al., 2017), steganography (Karakiş et al., 2015) and so on. Kokare et al. (Kokare et al., 2003) compared Euclidean, Manhattan, Chebychev, Mahalanobis, Canberra, Bray-Curtis, Weighted-Mean-Variance, Squared Chord, and Squared Chi-Squared distances/similarity function for measuring the texture retrieval from image. Puzicha et al. (Puzicha et al., 1997) presented a similarity measure based on non-parametric statistical tests to compare the empirical distributions of Gabor coefficients for textures in images. Holden et al. (2000) evaluated different similarity measure functions for rigid body registration of serial magnetic resonance brain scans. Two probabilistic similarity measures were proposed by Aksoy et al. (Aksoy and Haralick, 2001) and compared with geometric similarity measures for image retrieval. Wang and Simoncelli (2005) proposed the similarity measure that works in the wavelet frequency domain of image to measure the translation, scaling and rotation of images.

A similarity assessment technique using fuzzy logic (FL) for judgment of properties was proposed by Santini et al. (Santini and Jain, 1999) to feature contrast in images. To distinguish meaningful objects in images, Chien et al. (Chien and Cheng, 2002) developed an image segmentation scheme that was based on fuzzy color similarity measure. A comparative study of crisp and fuzzy logic based similarity measures was done by Jain et al. (1995) to find the similar or distinct textures of the images. To handle possible distortions in fingerprints of a non-linear nature, Chen et al. (Chen et al., 2006) developed a normalized fuzzy similarity measure.

A fuzzy diffusion technique was proposed by Elmas et al. (2013) using the FL to obtain the similarity between pixels and generate a similar image used as a heat diffusion coefficient. Demirci et al. (Demirci, 2007) used similarity in edge detection in images. The authors used the pixel window to get the similarity value of the central pixel to its neighboring pixels and created a similarity matrix to detect edges in the image by using the threshold value. Wu et al. (Wu and Tsai, 2003) proposed a steganographic method in which secret data was embedded in those pixels where the pixel-value differences of two consecutive pixels were low. An adjacent pixel difference based steganographic technique was presented by Li et al. (Y.-C. Li et al., 2010) in which the histogram of adjacent pixels was employed to increase the capacity of embedding. Karakis et al. (Karakiş et al., 2015) proposed rule-based FL for image steganographic scheme to hide patient data in medical images.

## 3. Preliminaries

This section discusses the basics of interval type-2 fuzzy logic systems (IT2 FLS). In order to discuss an IT2 FLS, we must first define type-2 fuzzy sets (T2 FSs) and interval type-2 fuzzy sets (IT2 FSs).

### 3.1. T2 FSs and IT2 FSs

A T2 FS $\tilde{A}$, is distinguished by a membership function $\mu_{\tilde{A}}(x, u)$, defined over $x$ and $u$, where $x$ is an element taken from universe of discourse $X$ and $u$ represents its primary membership function, respectively (R. John, 1998). Thus, the T2 FS $\tilde{A}$ can be expressed as Eq. (1):

$$\tilde{A} = \{(x, u), \mu_{\tilde{A}}(x, u) | \forall x \in X, \forall u \in J_x \subseteq [0, 1]\} \quad (1)$$

If the secondary membership function $\mu_{\tilde{A}}(x, u) = 1$, then a T2 FS becomes an IT2 FS (Mendel and Wu, 2010) as defined in Eq. (2).

$$\tilde{A} = \{(x, u), \mu_{\tilde{A}}(x, u) = 1 | \forall x \in X, \forall u \in J_x \subseteq [0, 1]\} \quad (2)$$

The graphical representation of an IT2 FS can be seen in Figure 1. In Figure 1, lower membership function (LMF) and upper membership function (UMF) are denoted by $\underline{\mu}_{\tilde{A}}(x)$ and $\overline{\mu}_{\tilde{A}}(x)$, repectively. The area that lies between the LMF and UMF is known as the Footprint of Uncertainty (FOU), given be Eq. (3).

$$FOU = [\underline{\mu}_{\tilde{A}}(x), \overline{\mu}_{\tilde{A}}(x)] \quad (3)$$

### 3.2. Interval type-2 fuzzy logic systems (IT2 FLS)

An IT2 FLS consists of fuzzification, rule-based inference engine, type-reduction and defuzzification procedures. A typical IT2 FLS can be seen in diagram form in Figure 2 (Mendel and Wu, 2010). The fuzzification takes input as a crisp value and consequently generates an IT2 FS using interval type-2 membership functions (IT2 MF). By applying certain *if-then* rules, the inference engine combines all the input IT2 FSs and gives an output IT2 FS. The type-reduction process takes the output IT2 FS and reduces it to a T1 FS. After that, the type-reduced set is put through defuzzification to achieve a final crisp output.

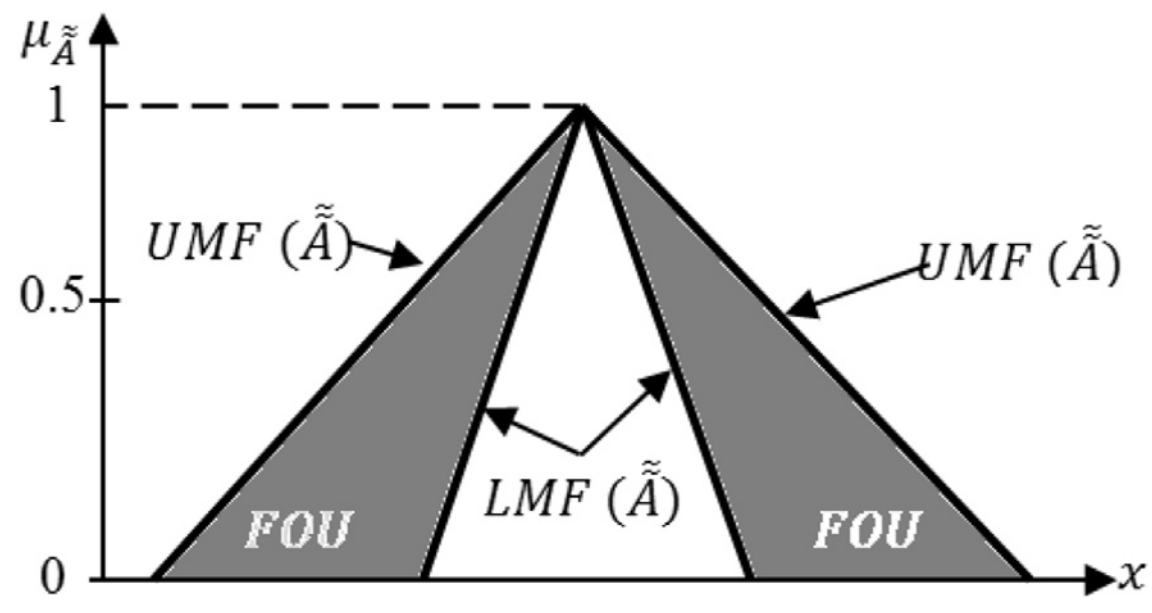

**Figure 1.** Interval type-2 fuzzy set.

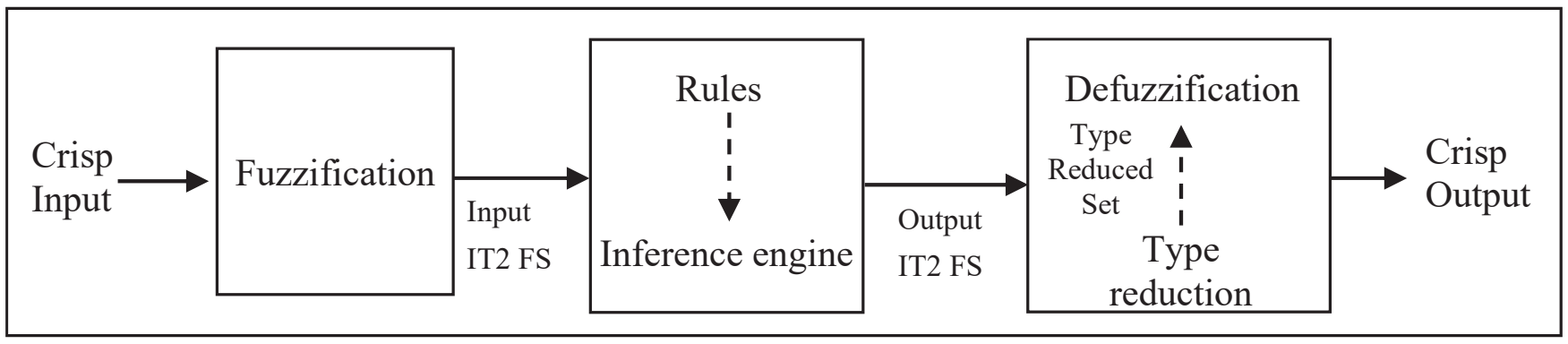

**Figure 2.** Interval type-2 fuzzy logic system.

## 4. Proposed steganographic method

The proposed steganography method named 'IT2 FLS based similarity measure for image steganography' is depicted in the form of a flow diagram in Figure 3. The detailed explanation of each step involved in the proposed method is discussed in the following sub-sections.

**Algorithm 1.** IT2 FLS based similarity measure

```
Input: Gray level image of size (H × W × 3)
Output: Similarity matrix (S_M)
For each H
  For each W
    Neighboring pixels of window (3 × 3)
    For each window
      Color differences (δR, δG, δB)      // Eqs. (4)-(6)
      Assign IT2 MFs                      // Fuzzification
      Inference-engine                    // Using Rules
      Type-Reduction                      // EKM Algorithm
      Defuzzification
    End
    Calculate S_a and S_M
  End
End
```

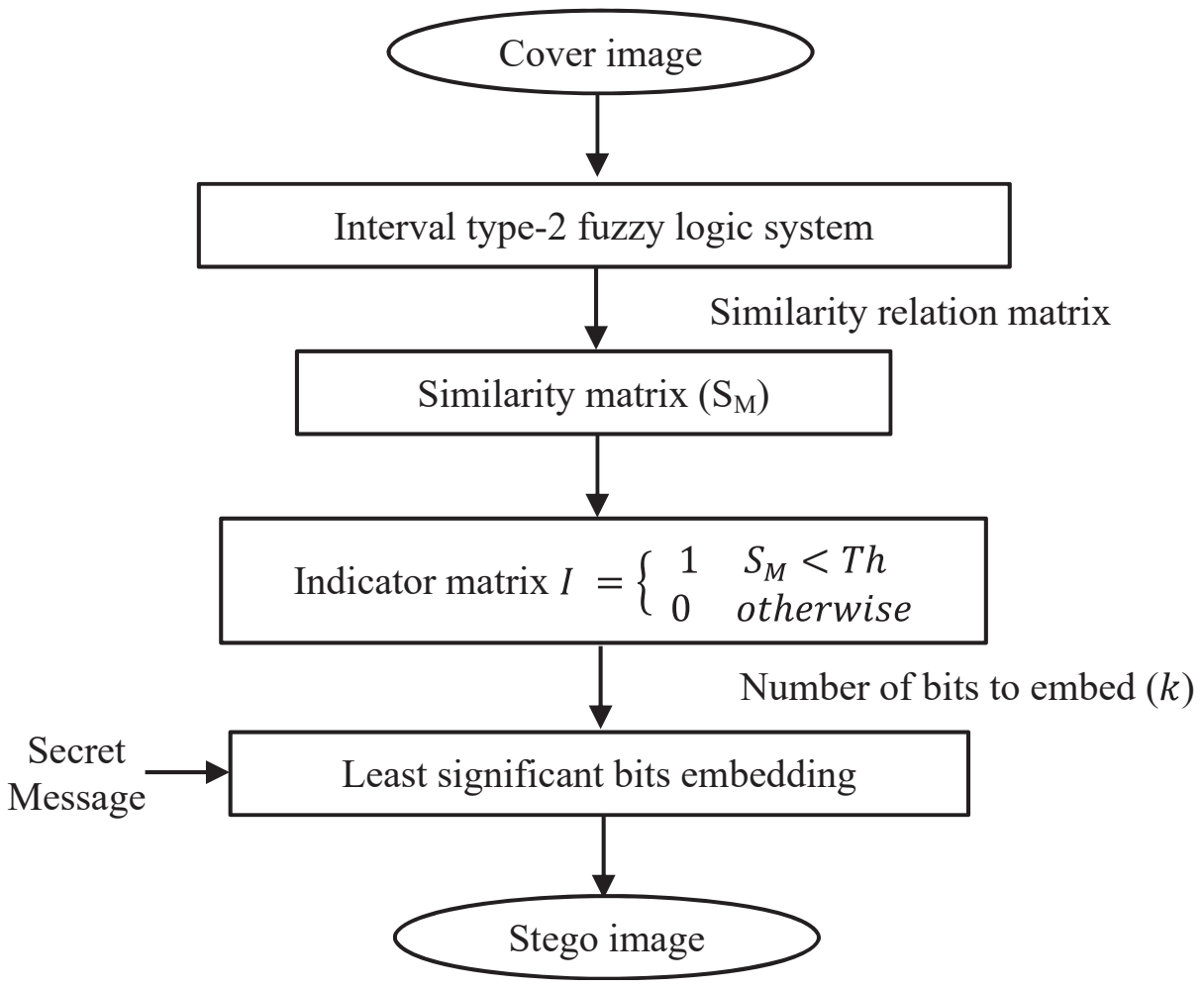

**Figure 3.** Proposed LSB image steganographic method using IT2 FLS based similarity measure.

### 4.1. IT2 FLS based similarity measure procedure

An image is comprised of pixels that are used to hide the secret information. A particular pixel is chosen for hiding if the value of similarity measure between the neighboring pixels is high. Figure 4 shows a typical window of size $(3 \times 3)$ of the image that illustrates the neighboring pixels $(P_1, P_2, \ldots, P_8)$ of the central pixel $P_9$. Therefore, pixel $P_9$ is selected for embedding if its similarity to its neighboring pixels is high.

In the IT2 FLS based similarity measure procedure, a $(3 \times 3)$ window corresponding to the each pixel in the cover image of size $(H \times W \times 3)$ is taken to calculate the similarity value of the central pixel. Algorithm 1 gives the step-by-step procedure of IT2 FLS based similarity measure. It consists of five main steps: (i) calculating gray level differences of pixels, (ii) interval type-2 fuzzification, (iii) rule-based inference engine, (iv) type-reduction and defuzzification, and (v) calculation of the similarity matrix. A detailed discussion on these steps is given in this sub-section. The output of Algorithm-1 is a similarity matrix $(S_M)$ of size $(H \times W)$, whose values lie within the interval $[0, 1]$. Each element belonging to the similarity matrix gives the similarity of the central pixel to its neighboring pixels in the window.

#### 4.1.1. Calculation of the gray-level differences

The pixels have three color components, Red $(\mathcal{R})$, Blue $(B)$, and Green $(G)$, with their corresponding gray-level intensities, $L_{\mathcal{R}}$, $L_B$ and $L_G$, respectively as given in Figure 5. The gray-level differences $\{\delta\mathcal{R}, \delta G, \delta B\}$ between two pixels $P_i$ and $P_j$ corresponding to each of the color components, $\{\mathcal{R}, G, B\}$, are calculated as follows:

$$\delta\mathcal{R} = |L_{\mathcal{R},i} - L_{\mathcal{R},j}| \tag{4}$$

$$\delta G = |L_{G,i} - L_{G,j}| \tag{5}$$

$$\delta B = |L_{B,i} - L_{B,j}| \tag{6}$$

#### 4.1.2. Interval type-2 fuzzification

In this step, the gray-level differences of each pair of pixels in all three color components $\{\delta\mathcal{R}, \delta G, \delta B\}$ are assigned interval type-2 membership functions (IT2 MFs). These IT2 MFs represent the corresponding linguistic terms: Low, Medium and High. There are several IT2 MFs such as triangular, trapezoidal, Gaussian, etc., available for handling these linguistic terms. We have used the triangular MFs to model them, because each of the linguistic values are provided in the form of intervals and their left and right ends are specific to the experts (Muhuri et al., 2017). The interval type-2 triangular membership function (IT2 TMF) is given in Eq. (7) which is defined using the LMF and the UMF given as Eqs. (8) and (9) respectively.

| | | |
|---|---|---|
| $P_1$ | $P_2$ | $P_3$ |
| $P_4$ | $\boldsymbol{P_9}$ | $P_5$ |
| $P_6$ | $P_7$ | $P_8$ |

**Figure 4.** Neighboring pixels.

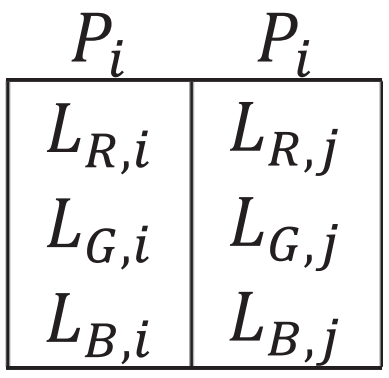

**Figure 5.** Gray levels of pixels.

$$\mu_{\tilde{A}_i}(x) = \left[\underline{\mu}_{\tilde{A}_i}(x), \overline{\mu}_{\tilde{A}_i}(x)\right], \forall x \in X \tag{7}$$

where

$$\underline{\mu}_{\tilde{A}_i}(x; \alpha, \beta, \gamma) = \begin{cases} 0, & x \le \alpha_i \\ \dfrac{x - \alpha_i}{\beta_i - \alpha_i}, & \alpha_i \le x \le \beta_i \\ \dfrac{\gamma_i - x}{\gamma_i - \beta_i}, & \beta_i \le x \le \gamma_i \\ 0, & x \ge \gamma_i \end{cases} \tag{8}$$

and

$$\overline{\mu}_{\tilde{A}_i}(x; \alpha', \beta, \gamma') = \begin{cases} 0, & x \le \alpha_i' \\ \dfrac{x - \alpha_i'}{\beta_i - \alpha_i'}, & \alpha_i' \le x \le \beta_i \\ \dfrac{\gamma_i' - x}{\gamma_i' - \beta_i}, & \beta_i \le x \le \gamma_i' \\ 0, & x \ge \gamma_i' \end{cases} \tag{9}$$

In Eqs. (8) and (9), $\alpha, \alpha', \beta, \gamma, \gamma'$ are five input parameters where $\beta \in [\alpha,\ \gamma]$ and $\beta \in [\alpha',\ \gamma']$ are used in the LMF $\underline{\mu}_{\tilde{A}_i}(x; \alpha, \beta, \gamma)$ and the UMF $\overline{\mu}_{\tilde{A}_i}(x; \alpha', \beta, \gamma')$, respectively. When the gray-level differences $\beta_i = \{\delta R,\ \delta G,\ \delta B\}$ fall in the interval $[\alpha,\ \gamma] = \{[0,\ 85],\ [86,\ 170],\ [171,\ 255]\}$, then the linguistic terms Low ($L$), Medium ($M$) and High ($H$) will be assigned to them, respectively. Figure 6 gives a typical example of IT2 TMFs representing the linguistic terms $L$, $M$and $H$.

#### *4.1.3. Rules and inference engine*

In the inference engine, the generated IT2 MFs assigned to each gray-level difference of the three color components of a pair of pixels are combined according to certain rules. As a result, a new IT2 MF describing the similarity between each pair of pixels is generated by assigning the linguistic terms such as *NS*- Not Similar, *SS*- Slightly Similar, *MS*- Moderately Similar, *QS*- Quite Similar, and *ES*- Exactly Similar. Figure 7 shows the IT2 MFs with respect to each linguistic term to describe the similarity.

The fuzzy-rules are given in Table 1 (Elmas et al., 2013). In Table 1, there are twenty seven fuzzy-rules listed to elaborate all the possible

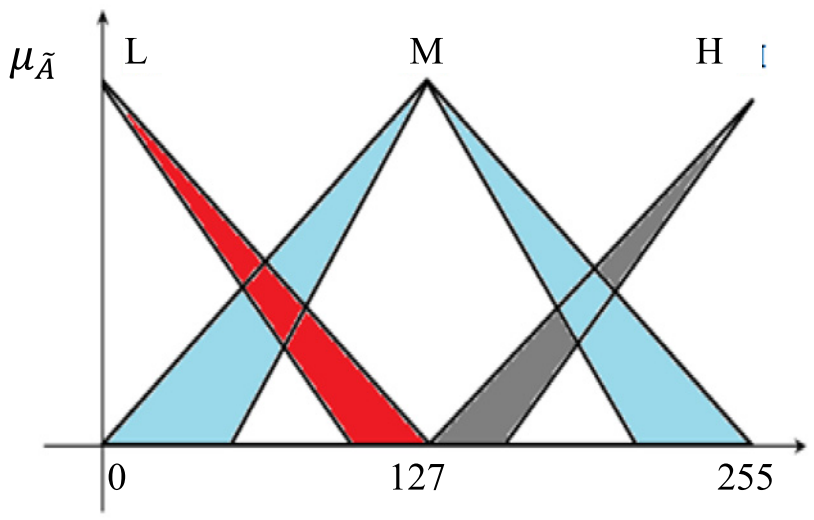

**Figure 6.** IT2 TMFs of the color component differences δR, δG, δB representing the linguistic terms.

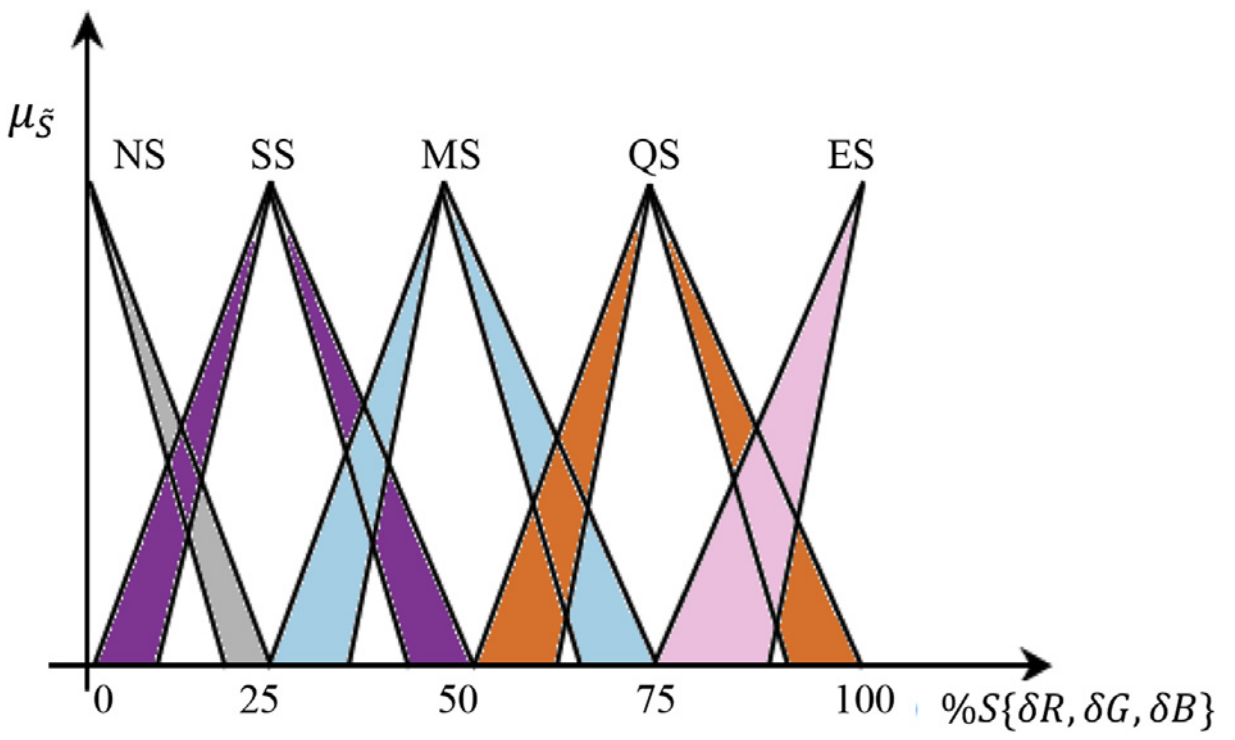

**Figure 7.** Typical IT2 TMFs of the similarities for the representation of the linguistic terms.

combinations of the input and output IT2 FSs. The inference engine maps the input difference IT2 FS to output similarity IT2 FS by combining rules. Let us consider that $x_1 \in X_1, \ldots, x_n \in X_n$ are $n$ inputs and $y \in Y$ is one output. Then, the $i^{\text{th}}$ rule given in the form of:

$R^i$ : *IF* $x_1$ *is* $\tilde{A}_1$, $x_2$ *is* $\tilde{A}_2 \ldots x_n$ *is* $\tilde{A}_n$ *Then* $y$ *is* $\tilde{B}$

can be written as Eq. (10)

$$R^l : = \tilde{A}_l \rightarrow \tilde{B}_l \quad l = 1, \ldots, M \tag{10}$$

where, $\tilde{A}_l = \tilde{A}_1^l \times \ldots \times \tilde{A}_P^l$ and $R^l$ is the fuzzy relation described by IT2 MF $\mu_{\tilde{A}_l \rightarrow \tilde{B}_l}(x, y)$ as Eq. (11)

$$\begin{aligned} \mu_{\tilde{A}_l \rightarrow \tilde{B}_l}(x, y) &= \mu_{\tilde{A}_1^l}(x_1) \sqcap \ldots \sqcap \mu_{\tilde{A}_p^l}(x_p) \sqcap \mu_{\tilde{B}^l}(y) \\ &= \left[\sqcap_{i=1}^{p} \mu_{\tilde{A}_i^l}(x_i)\right] \sqcap \mu_{\tilde{B}^l}(y) \end{aligned} \tag{11}$$

In the IT2 FLS, we used product $t$-norm to perform the intersection operation ($\sqcap$), such that the result of combining the input and antecedent are contained in the firing set given as Eq. (12):

$$F^l(x') = \left[\underline{f}^l(x'), \overline{f}^l(x')\right] \tag{12}$$

where we use Eq. (13) to compute.

$$\overline{f}^l(x') = \left[\sqcap_{i=1}^{p} \overline{\mu}_{\tilde{A}_i^l}(x_i)\right] \sqcap \overline{\mu}_{\tilde{B}^l}(y) \tag{13}$$

Eq. (14) gives the centroid of the IT2 FS.

$$\underline{f}^l(x') = \left[\sqcap_{i=1}^{p} \underline{\mu}_{\tilde{A}_i^l}(x_i)\right] \sqcap \underline{\mu}_{\tilde{B}^l}(y) \tag{14}$$

#### *4.1.4. Type-reduction and defuzzification*

The output IT2 FS from the inference engine is transformed into T1 FS and crisp value through the process of type-reduction and defuzzification, respectively. The method used here is known as the Enhanced Karnik-Mendel (EKM) algorithm for type-reduction. The complete procedure of EKM algorithm (Mendel and Wu, 2010) is explained in Algorithm 2. We have included the salient points of the Algorithms 2 (a-b) as below:

The centroid $C_{\tilde{A}}(x)$ is the collection of all embedding T1 FSs in the IT2 FS $\tilde{A}$ as given in Eq. (15).

$$C_{\tilde{A}}(x) = \{c_l(\tilde{A}), \ldots, c_r(\tilde{A})\} = [c_l(\tilde{A}), c_r(\tilde{A})] \tag{15}$$

**Table 1.** Fuzzy rules (Ashraf et al., 2018b).

| Rule | Antecedent | Consequent | |
|---|---|---|---|
| $R^1$: | If $\mu_{\delta R}$ is L, $\mu_{\delta G}$ is L, $\mu_{\delta B}$ is L | then | $\mu_S$ is ES |
| $R^2$: | If $\mu_{\delta R}$ is L, $\mu_{\delta G}$ is L, $\mu_{\delta B}$ is M | then | $\mu_S$ is ES |
| $R^3$: | If $\mu_{\delta R}$ is L, $\mu_{\delta G}$ is L, $\mu_{\delta B}$ is H | then | $\mu_S$ is QS |
| $R^4$: | If $\mu_{\delta R}$ is L, $\mu_{\delta G}$ is M, $\mu_{\delta B}$ is L | then | $\mu_S$ is ES |
| $R^5$: | If $\mu_{\delta R}$ is L, $\mu_{\delta G}$ is M, $\mu_{\delta B}$ is M | then | $\mu_S$ is QS |
| $R^6$: | If $\mu_{\delta R}$ is L, $\mu_{\delta G}$ is M, $\mu_{\delta B}$ is H | then | $\mu_S$ is MS |
| $R^7$: | If $\mu_{\delta R}$ is L, $\mu_{\delta G}$ is H, $\mu_{\delta B}$ is L | then | $\mu_S$ is QS |
| $R^8$: | If $\mu_{\delta R}$ is L, $\mu_{\delta G}$ is H, $\mu_{\delta B}$ is M | then | $\mu_S$ is MS |
| $R^9$: | If $\mu_{\delta R}$ is L, $\mu_{\delta G}$ is H, $\mu_{\delta B}$ is H | then | $\mu_S$ is SS |
| $R^{10}$: | If $\mu_{\delta R}$ is M, $\mu_{\delta G}$ is L, $\mu_{\delta B}$ is L | then | $\mu_S$ is ES |
| $R^{11}$: | If $\mu_{\delta R}$ is M, $\mu_{\delta G}$ is L, $\mu_{\delta B}$ is M | then | $\mu_S$ is QS |
| $R^{12}$: | If $\mu_{\delta R}$ is M, $\mu_{\delta G}$ is L, $\mu_{\delta B}$ is H | then | $\mu_S$ is MS |
| $R^{13}$: | If $\mu_{\delta R}$ is M, $\mu_{\delta G}$ is M, $\mu_{\delta B}$ is L | then | $\mu_S$ is QS |
| $R^{14}$: | If $\mu_{\delta R}$ is M, $\mu_{\delta G}$ is M, $\mu_{\delta B}$ is M | then | $\mu_S$ is MS |
| $R^{15}$: | If $\mu_{\delta R}$ is M, $\mu_{\delta G}$ is M, $\mu_{\delta B}$ is H | then | $\mu_S$ is SS |
| $R^{16}$: | If $\mu_{\delta R}$ is M, $\mu_{\delta G}$ is H, $\mu_{\delta B}$ is L | then | $\mu_S$ is MS |
| $R^{17}$: | If $\mu_{\delta R}$ is M, $\mu_{\delta G}$ is H, $\mu_{\delta B}$ is M | then | $\mu_S$ is SS |
| $R^{18}$: | If $\mu_{\delta R}$ is M, $\mu_{\delta G}$ is H, $\mu_{\delta B}$ is H | then | $\mu_S$ is NS |
| $R^{19}$: | If $\mu_{\delta R}$ is H, $\mu_{\delta G}$ is L, $\mu_{\delta B}$ is L | then | $\mu_S$ is QS |
| $R^{20}$: | If $\mu_{\delta R}$ is H, $\mu_{\delta G}$ is L, $\mu_{\delta B}$ is M | then | $\mu_S$ is MS |
| $R^{21}$: | If $\mu_{\delta R}$ is H, $\mu_{\delta G}$ is L, $\mu_{\delta B}$ is H | then | $\mu_S$ is SS |
| $R^{22}$: | If $\mu_{\delta R}$ is H, $\mu_{\delta G}$ is M, $\mu_{\delta B}$ is L | then | $\mu_S$ is MS |
| $R^{23}$: | If $\mu_{\delta R}$ is H, $\mu_{\delta G}$ is M, $\mu_{\delta B}$ is M | then | $\mu_S$ is SS |
| $R^{24}$: | If $\mu_{\delta R}$ is H, $\mu_{\delta G}$ is M, $\mu_{\delta B}$ is H | then | $\mu_S$ is NS |
| $R^{25}$: | If $\mu_{\delta R}$ is H, $\mu_{\delta G}$ is H, $\mu_{\delta B}$ is L | then | $\mu_S$ is SS |
| $R^{26}$: | If $\mu_{\delta R}$ is H, $\mu_{\delta G}$ is H, $\mu_{\delta B}$ is M | then | $\mu_S$ is NS |
| $R^{27}$: | If $\mu_{\delta R}$ is H, $\mu_{\delta G}$ is H, $\mu_{\delta B}$ is H | then | $\mu_S$ is NS |

We apply the EKM algorithm to calculate the left and right endpoints of the interval, $c_l$ and $c_r$ respectively. The equations to calculate them are given below as Eq. (16) and Eq. (17):

$$c_l = \min_{\forall \theta_i \in [\underline{\mu},\ \overline{\mu}]} \left( \sum_{i=1}^{N} x_i \theta_i \Big/ \sum_{i=1}^{N} \theta_i \right) \tag{16}$$

$$c_r = \max_{\forall \theta_i \in [\underline{\mu},\ \overline{\mu}]} \left( \sum_{i=1}^{N} x_i \theta_i \Big/ \sum_{i=1}^{N} \theta_i \right) \tag{17}$$

**Algorithm 2(a).** Calculation for $c_l(L)$

| | |
|---|---|
| | Set $k = [N/2.4]$ (the nearest integer to $N/2.4$) and compute: |
| 1. | $a = \sum_{i=1}^{k} x_i \overline{\mu}_{\tilde{A}}(x_i) + \sum_{i=k+1}^{k} x_i \underline{\mu}_{\tilde{A}}(x_i)$ |
| | $b = \sum_{i=1}^{k} \overline{\mu}_{\tilde{A}}(x_i) + \sum_{i=k+1}^{k} \underline{\mu}_{\tilde{A}}(x_i)$ |
| 2. | Compute $c' = a/b$ |
| 3. | Find $k' \in [1, N-1]$ such that $x_{k'} \leq c' x_{k'+1}$ |
| 4. | Check if $k' = k$. If yes, stop and set $c' = c_l(L)$, and $k = L$. If no, go to step 5. |
| | Compute $s = \mathrm{sign}(k' - k)$ and |
| 5. | $a' = a + s \sum_{i=\min(k,k')+1}^{\max(k,k')} x_i [\overline{\mu}_{\tilde{A}}(x_i) - \underline{\mu}_{\tilde{A}}(x_i)]$. |
| | $b' = b + s \sum_{i=\min(k,k')+1}^{\max(k,k')} [\overline{\mu}_{\tilde{A}}(x_i) - \underline{\mu}_{\tilde{A}}(x_i)]$ |
| 6. | Compute $c''(k') = a'/b'$ |
| 7. | Set $c' = c''(k'), a = a', b = b'$ and go to Step 2. |

**Algorithm 2(b).** Calculation for $c_r(R)$

| | |
|---|---|
| | Set $k = [N/1.7]$ (the nearest integer to $N/1.7$) and compute: |
| 1. | $a = \sum_{i=1}^{k} x_i \underline{\mu}_{\tilde{A}}(x_i) + \sum_{i=k+1}^{k} x_i \overline{\mu}_{\tilde{A}}(x_i)$ |
| | $b = \sum_{i=1}^{k} \underline{\mu}_{\tilde{A}}(x_i) + \sum_{i=k+1}^{k} \overline{\mu}_{\tilde{A}}(x_i)$ |
| 2. | Compute $c' = a/b$ |
| 3. | Find $k' \in [1, N-1]$ such that $x_{k'} \leq c' x_{k'+1}$ |
| 4. | Check if $k' = k$. If yes, stop and set $c' = c_r(R)$, and $k = R$. If no, go to step 5. |
| | Compute $s = \mathrm{sign}(k' - k)$ and |
| 5. | $a' = a - s \sum_{i=\min(k,k')+1}^{\max(k,k')} x_i [\overline{\mu}_{\tilde{A}}(x_i) - \underline{\mu}_{\tilde{A}}(x_i)]$. |
| | $b' = b - s \sum_{i=\min(k,k')+1}^{\max(k,k')} [\overline{\mu}_{\tilde{A}}(x_i) - \underline{\mu}_{\tilde{A}}(x_i)]$ |
| 6. | Compute $c''(k') = a'/b'$ |
| 7. | Set $c' = c''(k'), a = a', b = b'$ and go to Step 2. |

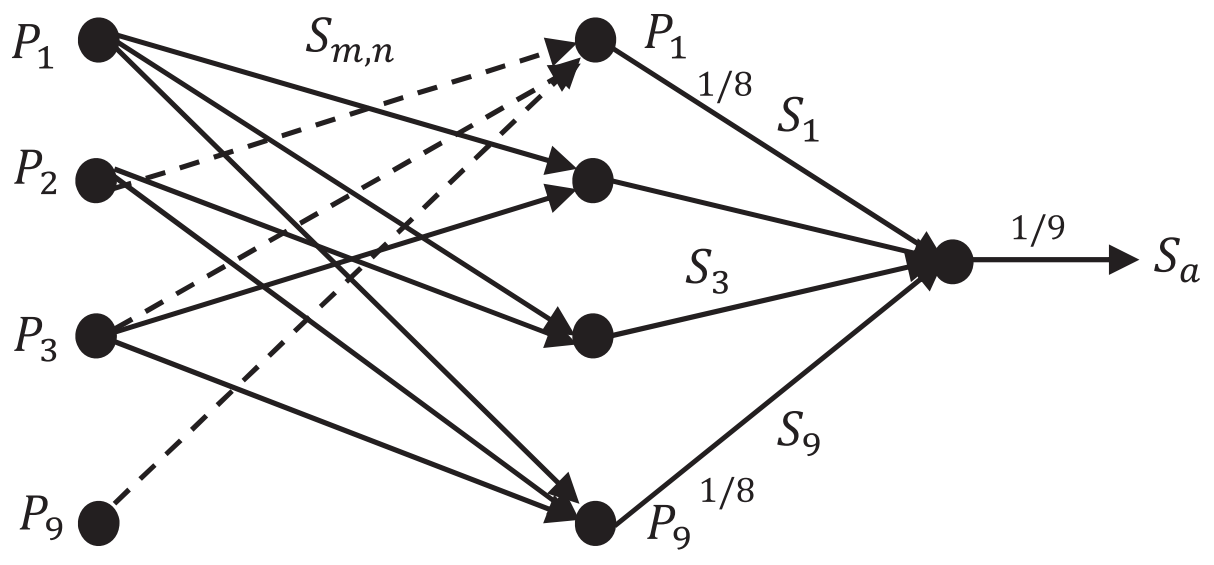

**Figure 8.** Similarity network[11] (Demirci, 2007).

Further, defuzzification produces a crisp output using the centroid method. This procedure of defuzzification uses the left centroid ($c_l$) and the right centroid ($c_r$) (Mendel and Wu, 2010). The final crisp output ($y_d$) is then obtained by taking a simple mean of $c_l$ and $c_r$, as given below in Eq. (18):

$$y_d = \frac{c_l + c_r}{2} \tag{18}$$

### 4.2. Similarity evaluation

The IT2 FLS thus calculates the similarity value of every pair of pixels in the $(3 \times 3)$ window, as may be seen in Figure 8 of the similarity network. The final result will be a similarity relation matrix that can be shown as

$$S_{m,n} = \begin{bmatrix} S_{1,1} & S_{1,2} & \cdots & S_{1,9} \\ S_{2,1} & S_{2,2} & \cdots & S_{2,9} \\ \vdots & \vdots & \ddots & \vdots \\ S_{9,1} & S_{9,2} & \ldots & S_{9,9} \end{bmatrix} \tag{19}$$

In Eq. (19), $S_{m,n}$ represents a similarity relation matrix, with all the elements belonging to the interval $[0, 1]$. Therefore, we can calculate the similarity of every $i^{th}$ pixel in the $(3 \times 3)$ window to its neighboring pixels as follows:

$$S_i = \frac{1}{8}\sum_{n=1}^{9} S_{i,n} \quad \text{for } i \neq n \tag{20}$$

To calculate the similarity of the central pixel in the window, we use Eq. (20) as follows:

$$S_a = \frac{1}{9}\sum_{n=1}^{9} S_i. \tag{21}$$

In Eq. (21), $S_a$ is the similarity of a single pixel in the image to its immediate surrounding neighbor pixels. Hence, the whole process of IT2 FLS and calculation of $S_a$ is iterated for every pixel in the image. The final result of this will be a similarity matrix ($S_M$) that holds the similarity value $S_a$ corresponding to each pixel of the cover image.

### 4.3. Similarity threshold

In order to choose which pixel is used for embedding, we employ a similarity threshold ($Th$) to generate an indicator matrix ($I$), given as follows:

$$I_{i,j} = \begin{cases} 1, & S_{M_{i,j}} \geq Th \\ 0, & S_{M_{i,j}} < Th \end{cases} \tag{22}$$

where $Th$ is the similarity threshold. The indicator matrix is a matrix of size $(H \times W)$ whose values are either 1, if the corresponding value of the similarity matrix $S_M$ is higher than the threshold $Th$, or 0, if it less than $Th$.

**Algorithm 3.** Embedding procedure

**Input**: Cover image ($C$), indicator matrix ($I$), Number of bits ($k$), Secret message ($m$).
**Output**: Stego image
**For** each color level
  **For** each $H$   // height of cover image
    **For** each $W$   // width of cover image
      **If** $I == 1$
        Convert pixel value into binary bit stream
        Replace $k$ LSB bits with the $m$ bits
        Convert changed binary into new pixel value
        Assign new pixel value to stego
      **End**
    **End**
  **End**
**End**

**Algorithm 4.** Extracting procedure

**Input**: Stego image, Number of bits ($k$)
**Output**: Secret message ($m$).
**For** each color level
  **For** each $H$   // height of cover image
    **For** each $W$   // width of cover image
      **If** $I == 1$
        Convert pixel value into binary bit stream
        Extract $k$ LSB bits from binary pixel value
        Combine all the extracted bits into $m$
      **End**
    **End**
  **End**
**End**

### 4.4. Embedding procedure

In the embedding procedure of the proposed algorithm, we used the least significant bit (LSB) embedding procedure to hide the message. In the LSB method, the decimal pixel value is converted into the binary bit stream, and the last $k$ bits are then replaced with the bits of the secret message (Ker, 2005; Yang et al., 2008). The output of the LSB embedding procedure is a new bit stream which is then converted back into decimal. The pixel of the cover image is selected for embedding if its value is higher than the threshold. The whole procedure is repeated for every pixel of the cover image. The final output of the IT2 FLS-LSB steganographic method is the stego image. Algorithm 3 gives the complete embedding procedure.

### 4.5. Extracting procedure

In the extracting procedure, the indicator matrix ($I$) is used to find the location of the embedding pixels of the stego image. For every embedding pixel, the last $k$ bits from its LSB are extracted and combined. The output of this procedure is the hidden secret message. Algorithm 4 gives the complete extracting procedure.

## 5. Experimental simulations and comparative analysis

The experimental simulations are performed to show the integrity of the proposed IT2 FLS based similarity measure in image steganography. The experimental simulations are conducted in MATLAB and run on Intel(R) Xeon(R) processor with 16 GB of RAM (3.40 GHz, Windows 7, 64 bits). We have used a set of well-known images as dataset (Cheddad et al., 2010; Johnson and Katzenbeisser, 2000; Petitcolas et al., 1999). The perceptual transparency of the proposed algorithm is measured using different quality metrics. Since, the least significant bits (LSB) procedure is used to perform embedding (as discuss in Section 4.4), we have termed the proposed IT2 FLS based similarity measure for LSB embedding as IT2 FLS-LSB steganography scheme.

Moreover, we have considered two more methods for evaluating the similarity of a pixel in an image as: (i) type-1 fuzzy logic system (T1 FLS) based similarity measure (Karakiş et al., 2015) and (ii) Euclidean distance based similarity measure (Demirci, 2007). Utilizing these two procedures for LSB embedding, we have develop and implemented two more methods, referred as, T1 FLS based similarity measures for least significant bits (T1FLS-LSB) steganographic method and Euclidean distance based similarity measures for least significant bit (SM-LSB) steganographic method.

In T1 FLS-LSB method, the similarity relation matrix (as in Eq. (19)) is calculated by the T1 FLS where the MFs are the TI MFs. Mamdani-rules are used in the inference engine and centroid approach is applied to calculate the defuzzify similarity value of a $(3 \times 3)$ window. The obtained similarity relation matrix is used to evaluate the similarity of a pixel (as in Eq. (21)) and, hence the indicator matrix (Eq. (22)) is generated via the threshold value. Finally, the embedding procedure, using Algorithm 3, is applied to obtain the stego image. For the SM-LSB method the distance between any two pixels is calculated by Euclidean nom as $D_{i,j} = \frac{1}{3}\sqrt{(\delta\mathcal{R}^2 + \delta G^2 + \delta B^2)}$, where $\delta\mathcal{R}$, $\delta G$ and $\delta B$ are three color components calculated by Eqs. (4), (5), and (6). Using the distance matrix, the similarity relation matrix (as in Eq. (19)) is calculated and the similarity of a pixel is obtained. Thus, the LSB embedding procedure is applied using indicator matrix to to obtain the stego image. Both of these approaches are used to perform steganography and the obtained results are compared with our proposed IT2 FLS-LSB scheme.

In the following subsections, we have discussed about the image dataset which was taken to perform the experimental simulations and the quality matrices. All the three steganographic methods are applied on these images and the metrics are calculated. The obtained stego images and the results are shown and thoroughly compared. Finally, we have done a comparative analysis of the proposed approach with some existing and well-known steganographic methods.

### 5.1. Dataset and quality matrices

The performances of all three steganographic procedures evaluated by using ten standard gray-level cover images, namely Lena, Baboon, Jet, Barbara, Boat, Peppers, Earth from space, House, Sailboat, and Splash, each of size $(512 \times 512 \times 3)$ pixels, shown in Figure 9 (a)-(j). We chose the Lena image of size $(256 \times 256 \times 3)$ pixels shown in Figure 10, as the secret message.

To measure the perceptual transparency of the proposed steganographic procedure, we have used three quality matrices, namely, PSNR, SSIM, and UQI (Subhedar and Mankar, 2014). The mathematical formulae for calculating the aforementioned measures are given below:

The PSNR compares the cover and stego images to test the effect of embedding, defined as follows:

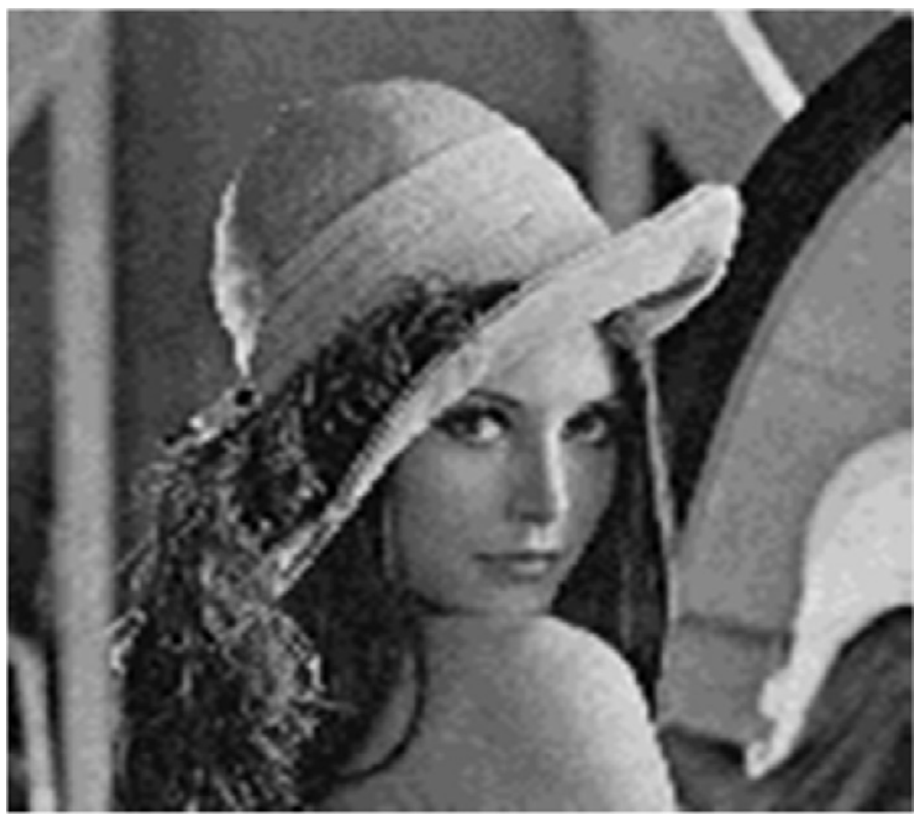

**Figure 10.** Secret Message: (Gray-scale Lena image of size $(256 \times 256 \times 3)$ pixels).

$$PSNR = 10 \log\left(\frac{255 \times 255}{MSE}\right) \tag{23}$$

In Eq. (23), MSE is the mean square error and it is calculated as follows using Eq. (24):

$$MSE = \frac{1}{MN}\sum_{i=0}^{M-1}\sum_{j=0}^{N-1}\left(f\left(x_i, y_j\right) - g\left(x_i, y_j\right)\right)^2 \tag{24}$$

where $M$ and $N$ are the dimensions of the images; $f(x,y)$ and $g(x,y)$ are the cover and the stego images respectively. The higher the values of PSNR, better the quality of the stego image. The lowermost PSNR value for an acceptable image steganographic method is 32 dB (Cheddad et al., 2010).

SSIM measures the similarity between the cover and the stego images. However, UQI measures the distortion between the cover and the stego images. They are defined as:

If $x = \{x_i | i = 1, ..., N\}$ is considered to be the original image and $y = \{y_i | i = 1, ..., N\}$ to be the stego image, then SSIM and the UQI are defined as Eq. (25) and Eq. (26), respectively:

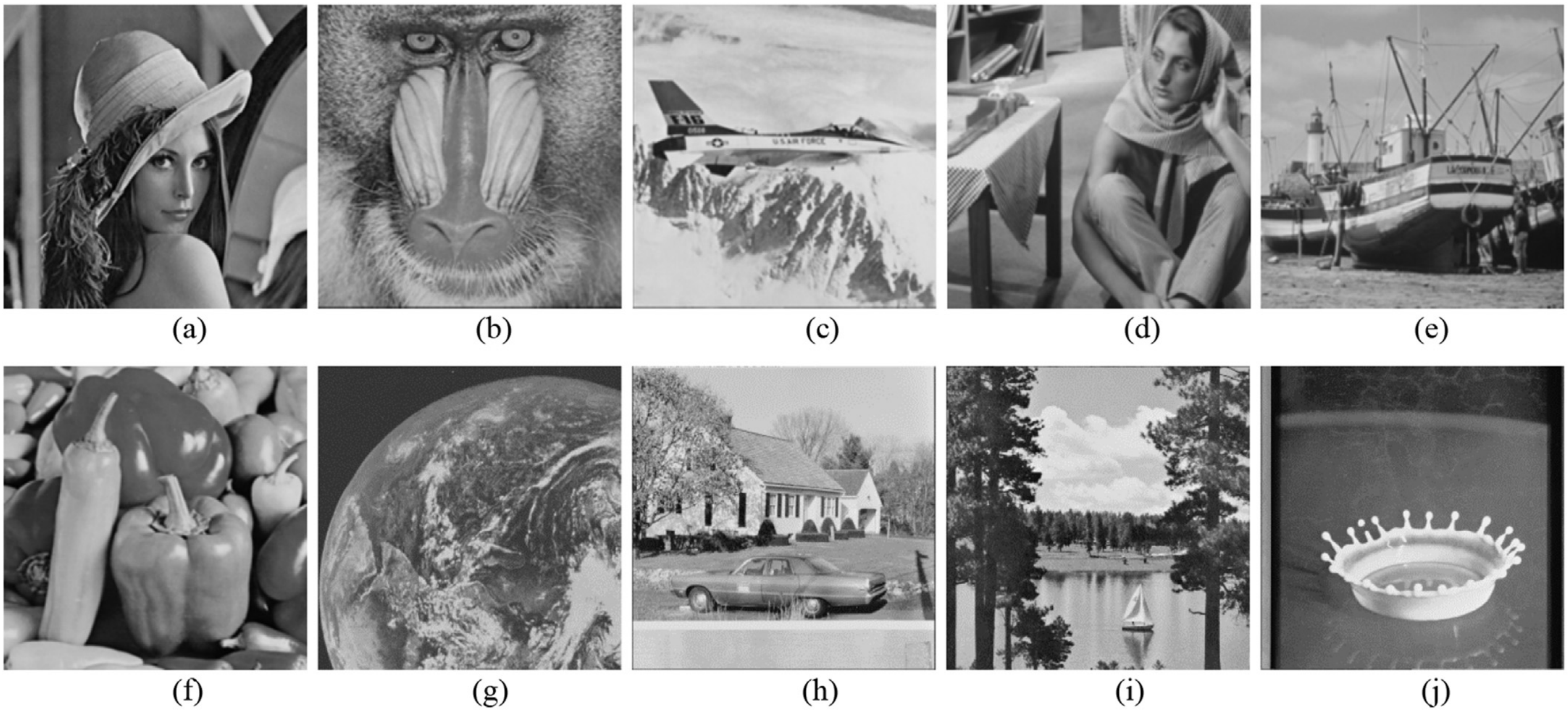

**Figure 9.** Cover images: (a) Lena, (b) Baboon, (c) Jet, (d) Barbara, (e) Boat, (f) Peppers, (g) Earth from space, (h) House, (i) Sailboat, (j) Splash.

$$SSIM(x,y) = \frac{(2\mu_x\mu_y + C_1)(2\sigma_{xy} + C_2)}{(\mu_x{}^2 + \mu_y{}^2 + C_1)(\sigma_x{}^2 + \sigma_y{}^2 + C_2)} \tag{25}$$

$$UQI(x,y) = \frac{4\sigma_{xy}.\mu_x.\mu_y}{(\sigma_x{}^2 + \sigma_y{}^2)[\mu_x{}^2 + \mu_y{}^2]} \tag{26}$$

where, $C_1, C_2, C_3$ are the constants; $\mu_x$ and $\mu_y$ are the mean, $\sigma_x{}^2$ and $\sigma_y{}^2$ are the variance of $x$ and $y$, and $\sigma_{xy}$ is the covariance between $x$ and $y$. The values of the SSIM and UQI measures lie within the range $[0,1]$, with 0 indicating poor quality, while 1 indicating excellent quality.

### 5.2. Results and comparisons

For better study of the proposed steganographic method, we took different number of bits ($k$) to embed into the LSBs of the selected pixels. Across all cover images, the number of bits are $k = \{1,2,3,4\}$. This means, after calculating the similarity matrix using the IT2 FLS based similarity measure and converting the secret message into a bit stream, we embed $k$ bits of the secret message into the $k$ LSBs of the selected pixels. The same process is replicated for T1 FLS-LSB and SM-LSB procedures for the number of bits.

In order to test the efficacy of the proposed steganographic method over T1 FLS-LSB and SM-LSB, we have chosen different thresholds ($Th$). As discussed previously, these thresholds are applied to the similarity matrices ($S_M$) to obtain the indicator matrices ($I$). The matrix $I$ indicates the location of the pixels selected for embedding. For each $Th$, the matrix $I$ is also embedded in the cover image along with the secret message to help in the extraction process. Since, $S_M$ contains values in the range $[0,1]$, we have arbitrarily chosen four different threshold values, $Th = \{0.75,\ 0.77,\ 0.80,\ 0.81\}$. The interpretation of $S_M > Th$ is that the similarity of that pixel is high, depicting a smooth or plain region of the image. Hence, that pixel is used for embedding. We perform the embedding procedure for each value of $k$, i.e., $k = \{1,2,3,4\}$ corresponding to all $Th = \{0.75,\ 0.77,\ 0.80,\ 0.81\}$ values using IT2 FLS-LSB, T1 FLS-LSB and SM-LSB steganographic methods and the stego images are generated.

The stego images for Lena, Baboon, Jet, Barbara, Boat, Peppers, Earth from space, House, Sailboat, and Splash cover images generated through the IT2 FLS-LSB for $k = 2$ and $Th = 0.80$ are depicted in Figure 11 (a)-(j). For $k = \{1,2,3,4\}$ and $Th = 0.80$, the quality metrics, i.e. PSNR, SSIM and UQI, comparing the visual transparency of the IT2 FLS-LSB, T1 FLS-LSB and SM-LSB steganographic methods for all the cover and stego images are shown in Table 2. For $k = 2$ and $Th = 0.80$, stego images for all cover images produced by the T1 FLS-LSB and SM-LSB steganographic methods are shown in Figures 12 (a)-(j) and 13 (a)-(j), respectively. From the Figures 11, 12, and 13, it can be observed that there are no much differences among the cover and stego images with respect to all three methods. A comparative plots, for $k = \{1,2,3,4\}$ and $Th = 0.80$, of the quality metrics, i.e. PSNR, SSIM and UQI, comparing the visual transparency between the IT2 FLS-LSB, T1 FLS-LSB and SM-LSB methods for all the cover and stego images are depicted in Figures 14, 15 and 16, respectively. From Figs. 14 (a)–(d), we can observe that the PSNR values for the proposed IT2 FLS-LSB method are significantly high in compare to T1 FLS-LSB and SM-LSB methods for all images. Moreover, the number of bits for hiding the secret data within the pixels of images increases the value of PSNR deceases for all images. Though, the minimum values of PSNR are moderate high (approximately $\geq 32$), the differences between the cover and stego images will be negligible by the human eyes (Cheddad et al., 2010). Similarly, from Figures 15 (a)–(d) and 16 (a)-(d), we can see that SSIM and UQI values for the proposed IT2 FLS-LSB method are better in compare to T1 FLS-LSB and SM-LSB methods for almost all images. From the Table 2 and the Figures 14, 15, and 16, we observe that the achieved values of PSNR, SSIM and UQI are relatively higher for our proposed method compared to the other two. Thus we can see that the proposed IT2 FLS-LSB for steganography outperforms both T1 FLS-LSB and SM-LSB methods.

Results of PSNR, SSIM and UQI values corresponding to $k = \{1,2,3,4\}$ and $Th = 0.75$, $Th = 0.77$ and $Th = 0.81$ for Lena, Baboon, Jet, Barbara, Boat, Peppers, Earth from space, House, Sailboat, and Splash cover images with IT2 FLS-LSB, T1 FLS-LSB and SM-LSB methods are given in Appendix A. The comparative plots for $k = \{1,2,3,4\}$ and $Th = 0.75$, $Th = 0.77$ and $Th = 0.81$ corresponding to each quality metrics, i.e. PSNR, SSIM and UQI, comparing the visual transparency between the IT2 FLS-LSB, T1 FLS-LSB and SM-LSB methods for Lena, Baboon, Jet, Barbara, Boat, Peppers, Earth from space, House, Sailboat, and Splash images are depicted in Appendix B.

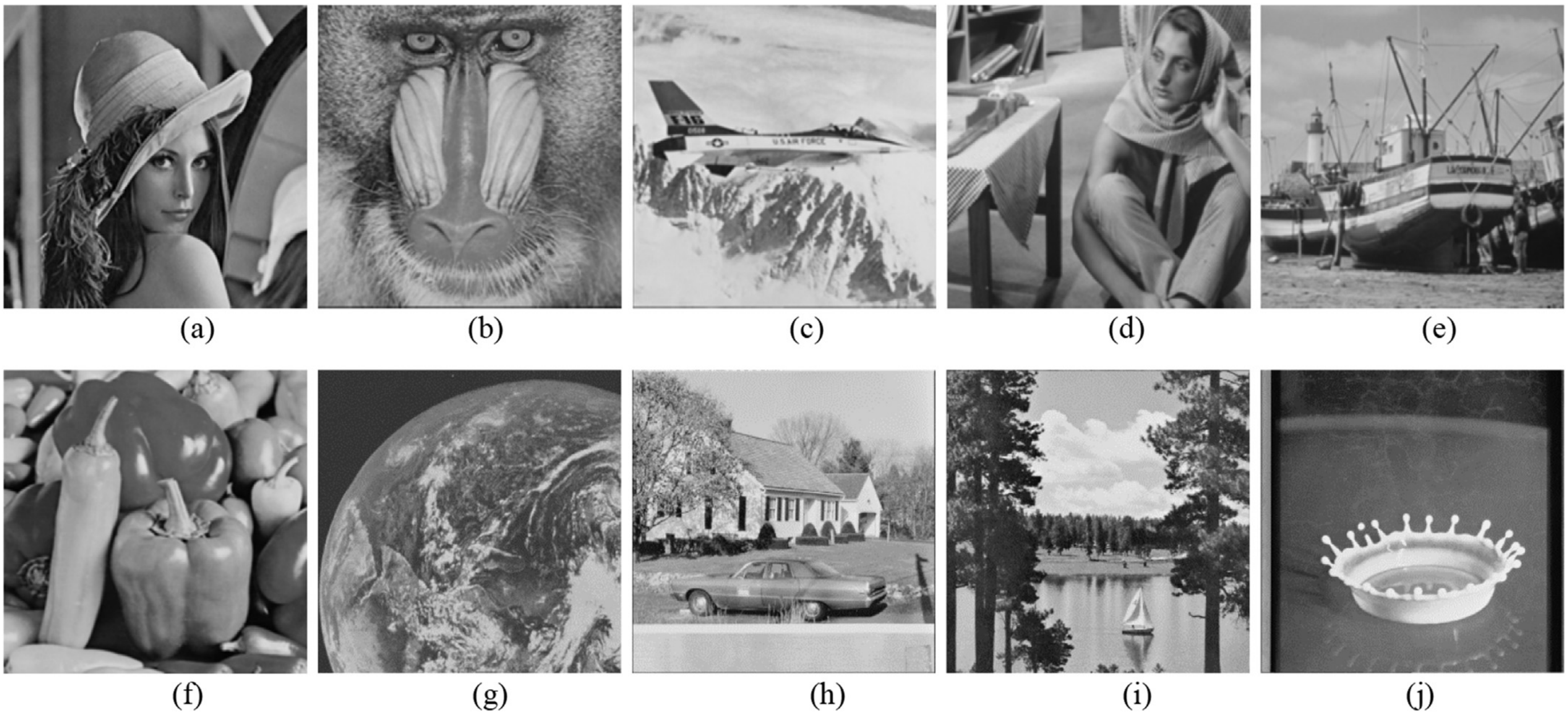

**Figure 11.** Stego images obtained via IT2 FLS-LSB scheme: (a) Lena, (b) Baboon, (c) Jet, (d) Barbara, (e) Boat, (f) Peppers, (g) Earth from space, (h) House, (i) Sailboat, (j) Splash.

**Table 2.** Quality matrices of SM-LSB, T1 FLS-LSB and proposed method for $k = \{1, 2, 3, 4\}$ and $Th = 0.80$

| | | SM-LSB | T1 FLS-LSB | Proposed | SM-LSB | T1 FLS -LSB | Proposed | SM-LSB | T1 FLS-LSB | Proposed | SM-LSB | T1 FLS-LSB | Proposed |
|---|---|---|---|---|---|---|---|---|---|---|---|---|---|
| Lena | PSNR | 51.2541 | 51.4005 | 51.6526 | 45.3338 | 45.4070 | 45.6694 | 39.1425 | 39.1561 | 39.4178 | 33.3102 | 33.5585 | 34.8785 |
| | SSIM | 0.9982 | 0.9964 | 0.9964 | 0.9937 | 0.9862 | 0.9863 | 0.9769 | 0.9486 | 0.9492 | 0.9276 | 0.8435 | 0.8451 |
| | UQI | 0.9999 | 0.9999 | 0.9999 | 0.9997 | 0.9995 | 0.9995 | 0.9987 | 0.9980 | 0.9981 | 0.9962 | 0.9926 | 0.9929 |
| Baboon | PSNR | 51.3801 | 52.1170 | 54.6426 | 45.3972 | 46.1367 | 48.6754 | 39.1431 | 39.8766 | 42.4214 | 33.3663 | 34.1130 | 36.6535 |
| | SSIM | 0.9987 | 0.9988 | 0.9989 | 0.9951 | 0.9953 | 0.9959 | 0.9814 | 0.9819 | 0.9844 | 0.9382 | 0.9400 | 0.9488 |
| | UQI | 0.9999 | 0.9999 | 0.9999 | 0.9995 | 0.9996 | 0.9998 | 0.9980 | 0.9982 | 0.9990 | 0.9924 | 0.9935 | 0.9962 |
| Jet | PSNR | 51.2764 | 51.4121 | 51.8790 | 45.2662 | 45.4127 | 45.8726 | 39.1003 | 39.2380 | 39.7138 | 33.2529 | 33.3899 | 33.8472 |
| | SSIM | 0.9955 | 0.9955 | 0.9956 | 0.9831 | 0.9831 | 0.9833 | 0.9404 | 0.9405 | 0.9413 | 0.8292 | 0.8292 | 0.8317 |
| | UQI | 0.9999 | 0.9999 | 0.9999 | 0.9995 | 0.9995 | 0.9996 | 0.9982 | 0.9982 | 0.9984 | 0.9933 | 0.9934 | 0.9940 |
| Barbara | PSNR | 51.3543 | 51.8192 | 52.9861 | 45.3624 | 45.8211 | 47.0036 | 39.1199 | 39.5811 | 40.7677 | 33.2855 | 33.7452 | 34.8956 |
| | SSIM | 0.9973 | 0.9973 | 0.9974 | 0.9896 | 0.9896 | 0.9901 | 0.9614 | 0.9617 | 0.9635 | 0.8830 | 0.8843 | 0.8902 |
| | UQI | 0.9999 | 0.9999 | 0.9999 | 0.9997 | 0.9997 | 0.9998 | 0.9988 | 0.9989 | 0.9991 | 0.9953 | 0.9958 | 0.9967 |
| Boat | PSNR | 51.2474 | 51.4245 | 52.1463 | 45.2171 | 45.3846 | 46.1282 | 39.0089 | 39.1952 | 39.9227 | 33.1830 | 33.3701 | 34.0806 |
| | SSIM | 0.9963 | 0.9963 | 0.9964 | 0.9858 | 0.9858 | 0.9862 | 0.9489 | 0.9492 | 0.9504 | 0.8529 | 0.8538 | 0.8578 |
| | UQI | 0.9999 | 0.9999 | 0.9999 | 0.9996 | 0.9997 | 0.9997 | 0.9986 | 0.9987 | 0.9989 | 0.9948 | 0.9950 | 0.9957 |
| Peppers | PSNR | 51.2661 | 51.3590 | 51.5642 | 45.1845 | 45.2665 | 45.4745 | 38.9167 | 38.9900 | 39.1970 | 33.0803 | 33.1559 | 33.3626 |
| | SSIM | 0.9961 | 0.9962 | 0.9962 | 0.9832 | 0.9833 | 0.9834 | 0.9408 | 0.9409 | 0.9415 | 0.8310 | 0.8312 | 0.8330 |
| | UQI | 0.9999 | 0.9999 | 0.9999 | 0.9996 | 0.9996 | 0.9996 | 0.9984 | 0.9985 | 0.9985 | 0.9941 | 0.9942 | 0.9944 |
| Earth | PSNR | 51.1830 | 51.3538 | 51.6727 | 45.1997 | 45.3733 | 45.7018 | 38.9573 | 39.1242 | 39.4430 | 33.2247 | 33.4066 | 33.7280 |
| | SSIM | 0.9973 | 0.9973 | 0.9973 | 0.9895 | 0.9896 | 0.9898 | 0.9606 | 0.9608 | 0.9614 | 0.8746 | 0.8757 | 0.8780 |
| | UQI | 0.9998 | 0.9998 | 0.9998 | 0.9993 | 0.9993 | 0.9993 | 0.9971 | 0.9972 | 0.9974 | 0.9893 | 0.9897 | 0.9904 |
| House | PSNR | 51.2704 | 51.6899 | 52.2922 | 45.3007 | 45.7154 | 46.3142 | 39.0425 | 39.4617 | 40.0706 | 33.2468 | 33.6588 | 34.2613 |
| | SSIM | 0.9968 | 0.9968 | 0.9969 | 0.9883 | 0.9885 | 0.9887 | 0.9580 | 0.9585 | 0.9594 | 0.8725 | 0.8744 | 0.8777 |
| | UQI | 0.9999 | 0.9999 | 0.9999 | 0.9997 | 0.9997 | 0.9997 | 0.9987 | 0.9988 | 0.9989 | 0.9950 | 0.9954 | 0.9959 |
| Sailboat | PSNR | 51.3154 | 51.6514 | 52.2377 | 45.3185 | 45.6644 | 46.2410 | 39.0950 | 39.4323 | 40.0174 | 33.3280 | 33.6613 | 34.2622 |
| | SSIM | 0.9973 | 0.9974 | 0.9974 | 0.9898 | 0.9901 | 0.9902 | 0.9624 | 0.9629 | 0.9638 | 0.8835 | 0.8852 | 0.8884 |
| | UQI | 0.9999 | 0.9999 | 0.9999 | 0.9997 | 0.9997 | 0.9998 | 0.9989 | 0.9990 | 0.9991 | 0.9959 | 0.9962 | 0.9966 |
| Splash | PSNR | 51.2545 | 51.2761 | 51.3652 | 45.1960 | 45.2269 | 45.3184 | 38.9479 | 38.9736 | 39.0675 | 33.1143 | 33.1340 | 33.2288 |
| | SSIM | 0.9941 | 0.9941 | 0.9942 | 0.9756 | 0.9756 | 0.9758 | 0.9176 | 0.9176 | 0.9186 | 0.7794 | 0.7790 | 0.7803 |
| | UQI | 0.9999 | 0.9999 | 0.9999 | 0.9997 | 0.9997 | 0.9997 | 0.9987 | 0.9987 | 0.9987 | 0.9949 | 0.9950 | 0.9951 |

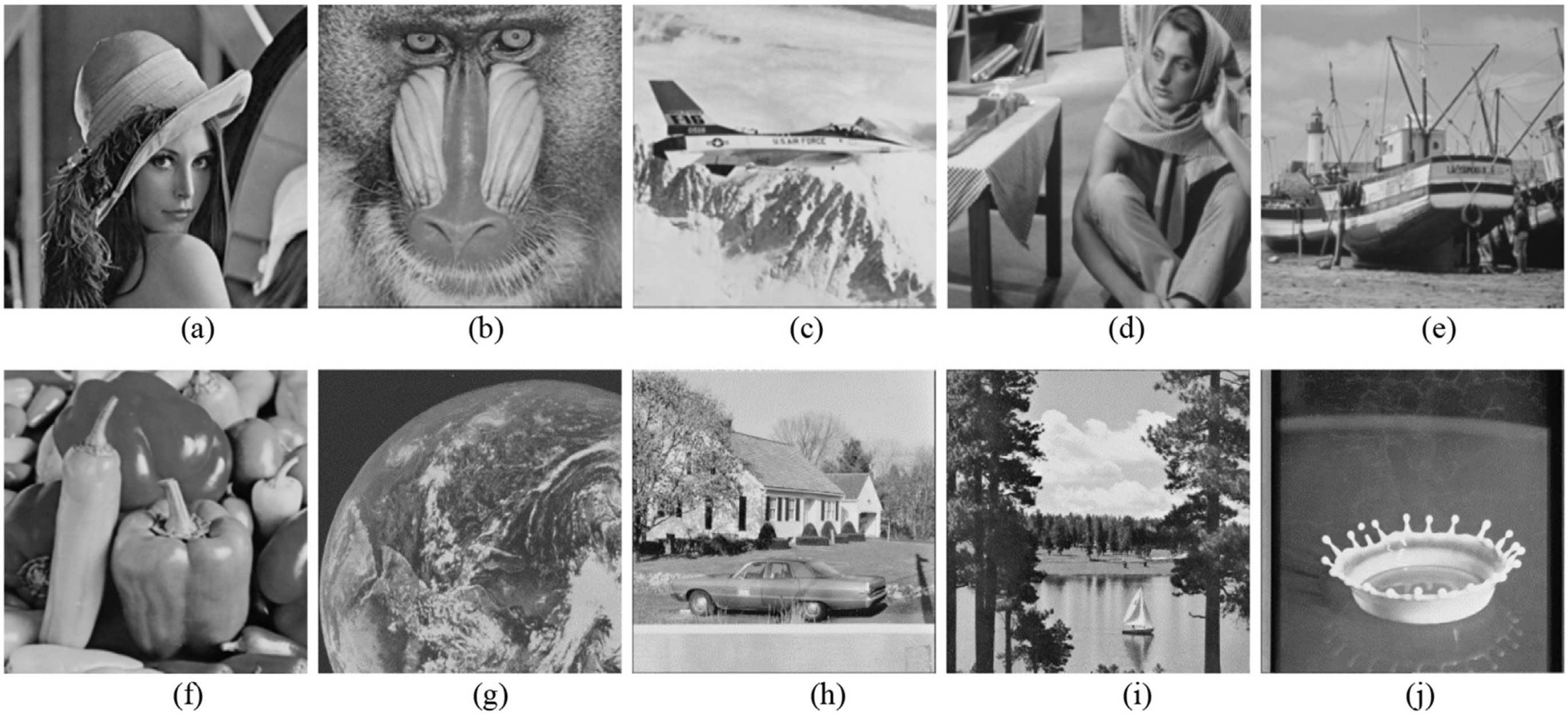

**Figure 12.** Stego images obtained via T1 FLS-LSB scheme: (a) Lena, (b) Baboon, (c) Jet, (d) Barbara, (e) Boat, (f) Peppers, (g) Earth from space, (h) House, (i) Sailboat, (j) Splash.

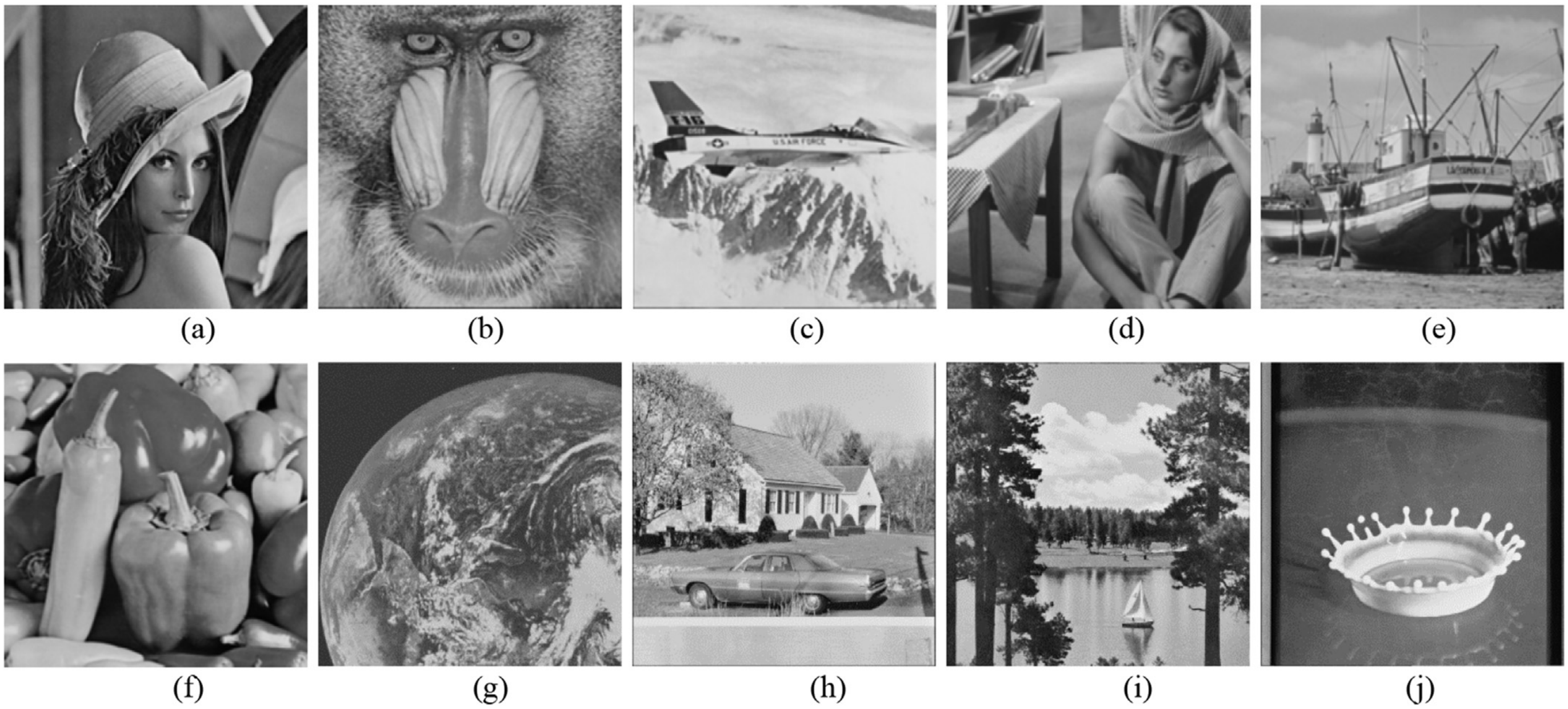

**Figure 13.** Stego images obtained via SM-LSB method: (a) Lena, (b) Baboon, (c) Jet, (d) Barbara, (e) Boat, (f) Peppers, (g) Earth from space, (h) House, (i) Sailboat, (j) Splash.

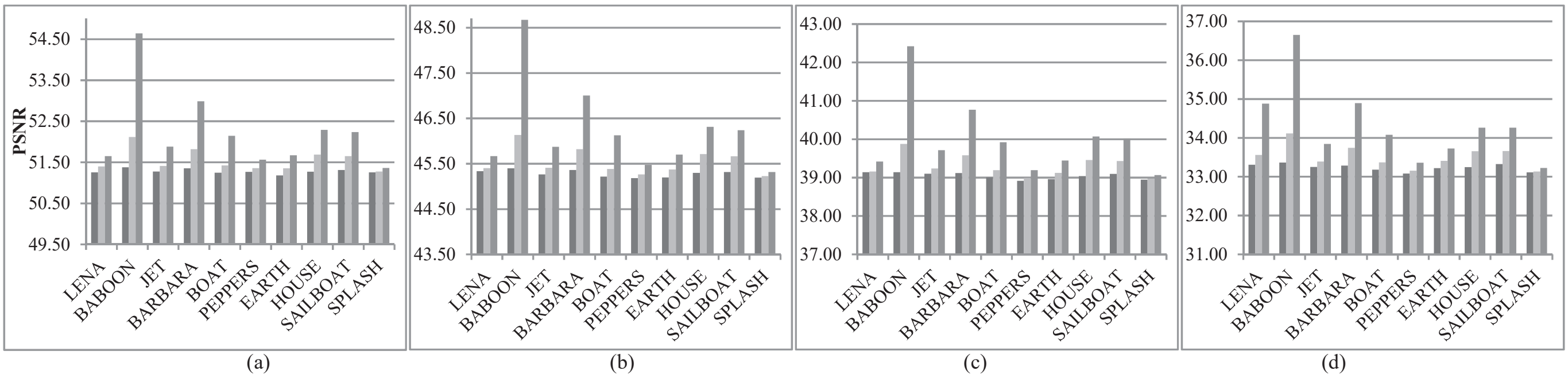

**Figure 14.** Comparison of PSNR of SM-LSB, T1 FLS-LSB and proposed methods for $Th = 0.80$: (a) $k = 1$, (b) $k = 2$, (c) $k = 3$, (d) $k = 4$ bits.

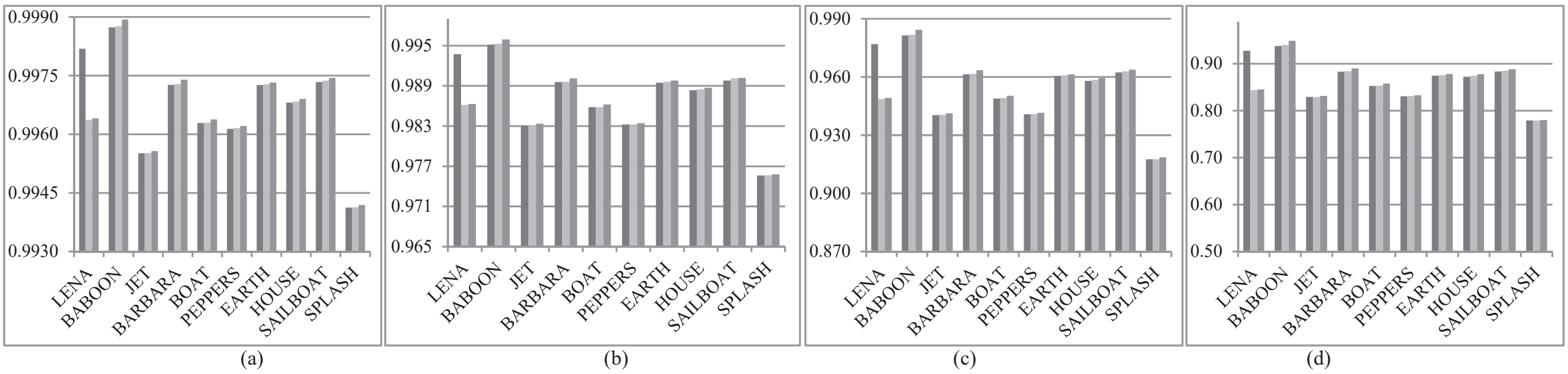

**Figure 15.** Comparison of SSIM of SM-LSB, T1 FLS-LSB and proposed methods for $Th = 0.80$: (a) $k = 1$, (b) $k = 2$, (c) $k = 3$, (d) $k = 4$ bits.

The embedding capacity can be defined as the amount of secret data that is hidden into an image. It is the ratio of number of bits hidden to the total number of bits. The embedding capacities (in percentage (%)) of the proposed IT2 FLS-LSB, T1 FLS-LSB and SM-LSB steganographic methods are presented in Table 3. In Table 3, the percentage embedding capacities for $k = \{1, 2, 3, 4\}$ corresponding to each threshold $Th = \{0.75, 0.77, 0.80, 0.81\}$ of the ten images used as the dataset are given. We have taken the average percentage embedding capacities across the ten images for each value of $k$ since, for each threshold, there are different amount of bits that are selected for embedding.

### 5.3. Comparison with other state of art stenographic schemes

In this section, we have compiled the similarity, differences, and some essential points of our proposed steganography scheme with other states of the art schemes. We have gone through many research works done in the past few decades on the image steganography. To the best of our knowledge, the IT2 FLS based steganographic method using the IT2 fuzzy rules and least significant bits (LSBs) has never appeared in the literature. For comparison, we have considered several research works that perform the embedding in the LSB domain of the cover image. Table 4 summaries the significant features of these prominent steganographic schemes (Amirtharajan and Balaguru Rayappan, 2012; Chan and Cheng, 2004; Sajasi and Eftekhari Moghadam, 2015; Wang et al., 2001), including the SM-LSB (Demirci, 2007), T1 FLS-LSB (Karakiş et al., 2015) and proposed IT 2FLS-LSB.

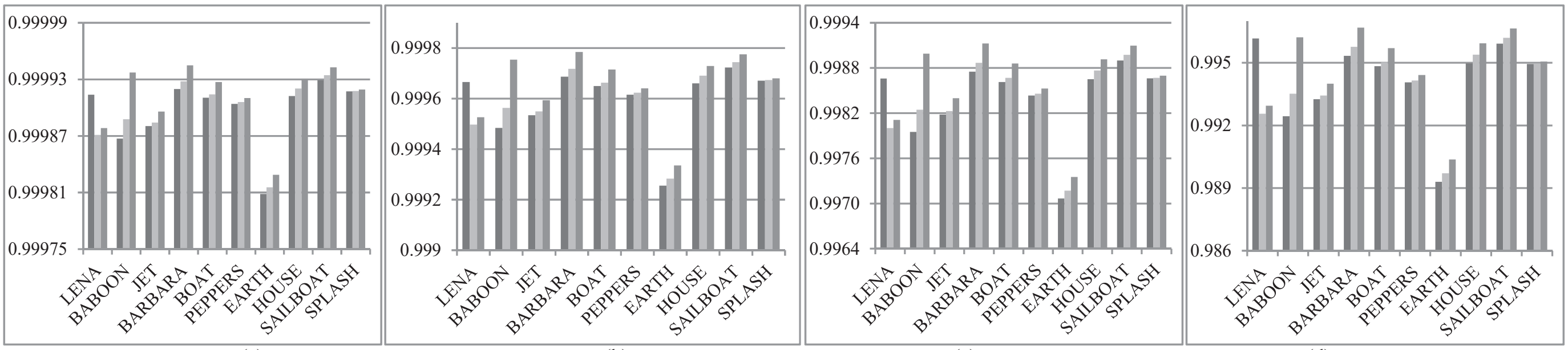

**Figure 16.** Comparison of UQI of SM-LSB, T1 FLS-LSB and proposed methods for $Th = 0.80$: (a) $k = 1$, (b) $k = 2$, (c) $k = 3$, (d) $k = 4$ bits.

**Table 3.** Embedding capacity in percentage (%).

| *Th* | SM-LSB | | | | T1 FLS-LSB | | | | IT2 FLS-LSB | | | |
|---|---|---|---|---|---|---|---|---|---|---|---|---|
| | $k = 1$ | $k = 2$ | $k = 3$ | $k = 4$ | $k = 1$ | $k = 2$ | $k = 3$ | $k = 4$ | $k = 1$ | $k = 2$ | $k = 3$ | $k = 4$ |
| 0.75 | 12.38 | 24.77 | 37.16 | 49.54 | 12.26 | 24.53 | 36.79 | 49.05 | 11.96 | 23.93 | 35.89 | 47.86 |
| 0.77 | 12.34 | 24.68 | 37.02 | 49.36 | 12.07 | 24.13 | 36.20 | 48.27 | 11.54 | 23.08 | 34.63 | 46.17 |
| 0.80 | 12.22 | 24.43 | 36.64 | 48.85 | 11.49 | 22.98 | 34.48 | 45.97 | 9.97 | 19.94 | 29.91 | 39.88 |
| 0.81 | 12.15 | 24.31 | 36.46 | 48.62 | 11.23 | 22.46 | 33.69 | 44.92 | 8.55 | 17.11 | 25.66 | 34.21 |

In Table 4, we have compared the hiding data size (bits), payload capacity (%) and PSNR (dB) for Lena, Baboon and Jet images that have been used in the different steganography schemes. A Genetic Algorithm (GA) based LSB steganographic scheme was developed by Wang et al. (2001). A secret data of 65,536 bits were embedded in cover images of size ($512 \times 512$). However, the payload capacity and PSNR values achieved by GA + LSB methods were inferior for all the Lena, Baboon and Jet images. Chan et al. (Chan and Cheng, 2004) proposed a steganographic scheme that performs the embedding by using the LSB method and utilized an optimal pixels adjustment procedure (OPAP) to reduce the distortion between the cover image and the stego image. The proposed approach was an extension of the simple LSB method (Wang et al., 2001) that could embed about 1,31, 608 bits and attain comparatively higher PSNR values (≅ 40.72) for the three images.

Amirtharajan et al. (Amirtharajan and Balaguru Rayappan, 2012) proposed adaptive LSB embedding approach. They considered cover images of size ($256 \times 256$), which was divided into blocks of ($4 \times 4$) pixels and encrypted secret messages were embedded in each block. For

**Table 4.** Comparative overview of the state of the art steganographic schemes.

| Cover | Studies | Method | Hiding data size (bits) | Payload Capacity (%) | PSNR (DB) |
|---|---|---|---|---|---|
| Lena | Wang et al. (Wang et al., 2001) | GA + LSB. | 65,536 | 25.00 | 38.72 |
| | Chan et al. (Chan and Cheng, 2004) | OPAP + LSB. | 1,31,072 | 50.00 | 40.72 |
| | Amritharajan et al. (Amirtharajan and Balaguru Rayappan, 2012) | Chaotic approach + LSB. | 65,536 | 25.00 | 38.66 |
| | Sajasi et al. (Sajasi and Eftekhari Moghadam, 2015) | NVF and Chaotic + LSB. | 81,920 | 31.25 | 44.48 |
| | Demirci (Demirci, 2007) | SM + LSB | 1,96,608 | 49.54 | 45.33 |
| | Karakis (Karakiş et al., 2015) | T1FLS + LSB. | 1,96,608 | 49.05 | 45.41 |
| | **Proposed scheme** | **IT2 FLS + LSB.** | 1,96,608 | 47.86 | 45.67 |
| Baboon | Wang et al. (Wang et al., 2001) | GA + LSB. | 65,536 | 0.25 | 38.73 |
| | Chan et al. (Chan and Cheng, 2004) | OPAP + LSB. | 1,31,072 | 50.00 | 40.72 |
| | Amritharajan et al. (Amirtharajan and Balaguru Rayappan, 2012) | Chaotic approach + LSB. | 65,536 | 25.00 | 38.67 |
| | Sajasi et al. (Sajasi and Eftekhari Moghadam, 2015) | NVF and Chaotic + LSB. | 81,920 | 31.25 | 47.66 |
| | Demirci (Demirci, 2007) | SM + LSB | 1,96,608 | 49.54 | 45.40 |
| | Karakis (Karakiş et al., 2015) | T1FLS + LSB. | 1,96,608 | 49.05 | 46.14 |
| | **Proposed scheme** | **IT2 FLS + LSB.** | 1,96,608 | 47.86 | 48.68 |
| Jet | Wang et al. (Wang et al., 2001) | GA + LSB. | 65,536 | 0.25 | 38.72 |
| | Chan et al. (Chan and Cheng, 2004) | OPAP + LSB. | 1,31,072 | 50.00 | 40.72 |
| | Amritharajan et al. (Amirtharajan and Balaguru Rayappan, 2012) | Chaotic approach + LSB. | 65,536 | 25.00 | 38.68 |
| | Demirci (Demirci, 2007) | SM + LSB | 1,96,608 | 49.54 | 45.27 |
| | Karakis (Karakiş et al., 2015) | T1FLS + LSB. | 1,96,608 | 49.05 | 45.41 |
| | **Proposed scheme** | **IT2 FLS + LSB.** | 1,96,608 | 47.86 | 45.87 |

k=3 bits, this scheme could insert about 65,536 bits and gave lower PSNR values of 38.66 dB, 38.67 dB and 38.68 dB corresponding to Lena, Baboon and Jet cover images. Sajasi et al. (Sajasi and Eftekhari Moghadam, 2015) proposed method produced high quality stego images PSNR values of 44.48 dB and 47.66 dB corresponding to Lena and Baboon, respectively though the payload capacity was 31.25%.

For similarity measure based least significant bit (SM + LSB) method (Demirci, 2007), type-1 fuzzy logic based least significant bit (T1 FLS + LSB) method (Karakiş et al., 2015) and the proposed interval type-1 fuzzy logic based least significant bit (IT2 FLS + LSB) method, the secret data size that are embedded in the cover images are much higher in comparison to others. The PSNR values for all three images are highest for the proposed IT2 FLS + LSB method in contrast to other approaches.

## 6. Conclusion

This work proposed an interval type-2 fuzzy logic system (IT2 FLS) based least significant bit (LSB) steganographic method, termed as IT2 FLS-LSB method, to perform steganography in high similarity valued pixels of an image. The IT2 FLS assigns interval type-2 membership functions (IT2 MFs) to color differences between pixels, expressed as linguistic variables such as low, medium and high. The idea of going into higher order fuzzy logic is to achieve better models of uncertainty and in this case they can be applied in image steganographic systems. Based on a set of interval type-2 fuzzy rules, the Mamdani inference engine operates on the IT2 MFs to produce the interval type-2 fuzzy similarity between the pixels in the images. To show the efficacy of our proposed method, we have also developed type-1 fuzzy logic system (T1 FLS)-LSB and similarity measure (SM)-LSB steganographic methods which evaluate pixel similarity through TI FLS and Euclidean distance, respectively. Based on a set of threshold values, which discriminate the similarity values of a pixel as high for these three steganographic methods, depicting a smooth or plain region of the image, we have embedded for different numbers of bits ($k = \{1, 2, 3, 4\}$) in the LSBs of the pixels. The experiments are performed on a dataset consisting of ten cover images, namely Lena, Baboon, Jet, Barbara, Boat, Peppers, Earth from space, House, Sailboat, and Splash, and the corresponding stego images are obtained. The perceptual transparency and visual quality of the stego images were measured by using three quality index metrics: peak signal-to-noise ratio (PSNR), structural similarity measure (SSIM), and universal quality index (UQI). We observed that the proposed IT2 FLS-LSB steganographic method outperforms both T1 FLS-LSB and SM-LSB methods, as the achieved values of PSNR, SSIM and UQI are relatively higher while maintaining comparable payload capacities.

Our future direction will be the extension of fuzzy bounded variation approach for the other higher order fuzzy sets like general type-2 fuzzy sets, interval type-2 fuzzy sets, and interval-valued-intuitionistic fuzzy sets, etc. Further, we will perform steganalysis on the proposed steganographic approach through the imperceptibility and robustness against different attacks. Moreover, instead of performing embedding in the spatial domain of the image, we may use transform domain techniques such as discrete wavelet transform, integer wavelet transform, and so on.

## Declarations

### *Author contribution statement*

Zubair Ashraf: Conceived and designed the experiments; Performed the experiments; Analyzed and interpreted the data; Contributed materials, analysis tools or data.

Mukul Lata Roy: Conceived and designed the experiments; Performed the experiments; Analyzed and interpreted the data; Wrote the paper.

Pranab K. Muhuri: Conceived and designed the experiments; Analyzed and interpreted the data; Contributed materials, analysis tools or data.

Q. M. Danish Lohani: Conceived and designed the experiments; Analyzed and interpreted the data.

### *Funding statement*

This research did not receive any specific grant from funding agencies in the public, commercial, or not-for-profit sectors.

### *Competing interest statement*

The authors declare no conflict of interest.

### *Additional information*

Supplementary content related to this article has been published online at https://doi.org/10.1016/j.heliyon.2020.e03771.

## Acknowledgements

The authors would like to express their sincere thanks to the reviewers and the editors of the journal for their helpful comments and suggestions that helped in the improvement of the manuscript. Authors gratefully acknowledge the infrastructural and research facilities provided by the South Asian University, New Delhi through the Computational Intelligence lab of the Department of Computer Science while designing the experiments and conducting investigation.